\documentclass[1p,times]{elsarticle}

\usepackage{amsmath,amssymb,amsfonts}
\usepackage{mathtools}
\usepackage{graphicx}
\usepackage{booktabs}
\usepackage{array}
\usepackage{tabularx}
\newcolumntype{C}{>{\centering\arraybackslash}X}
\newcolumntype{L}{>{\raggedright\arraybackslash}X}
\usepackage{xcolor}
\usepackage{siunitx}
\usepackage[hidelinks,pdfusetitle=false]{hyperref}
\hypersetup{pdftitle={Implications of an analysis of the shear transformation of a zone of an oblate spheroid for dislocation theory, grain boundary sliding and superplasticity},pdfauthor={Albert Linda and K. A. Padmanabhan}}
\usepackage{microtype}

\graphicspath{{figures/}}
\biboptions{authoryear}

\newcommand{\OSTZ}{\textnormal{\textsc{ostz}}}
\newcommand{\dF}{E_{\mathrm{el}}}
\newcommand{\gz}{\gamma_0}
\newcommand{\beff}{b_\mathrm{eff}}

\newcommand{\half}{\tfrac{1}{2}}

\newcommand{\rev}[1]{#1}
\newcommand{\parthead}[1]{\par\bigskip\noindent{\large\bfseries #1}\par
  \nopagebreak\medskip\addcontentsline{toc}{section}{#1}}

\makeatletter
\def\ps@pprintTitle{\let\@oddhead\@empty\let\@evenhead\@empty
  \def\@oddfoot{\reset@font\hfil\thepage\hfil}\let\@evenfoot\@oddfoot}
\makeatother
\usepackage{longtable}
\newcommand{\g}{\gamma_0}
\newcommand{\eps}{\varepsilon_0}
\begin{document}
\begin{frontmatter}

\title{Implications of an analysis of the shear transformation of a zone
  of an oblate spheroid for dislocation theory, grain boundary sliding and
  superplasticity}

\author[iitk]{Albert~Linda\corref{cor}}
\ead{albert.hzbn@gmail.com}
\author[anna]{K.A.~Padmanabhan\corref{cor}}
\ead{ananthaster@gmail.com}

\cortext[cor]{Corresponding author.}
\affiliation[iitk]{organization={Department of Materials Science and
  Engineering, Indian Institute of Technology Kanpur},
  city={Kanpur}, postcode={208016}, country={India}}
\affiliation[anna]{organization={Professor of Eminence, Materials
  Science and Engineering Programme, Department of Mechanical
  Engineering, Anna University},
  city={Chennai}, postcode={600\,025}, country={India; formerly of IIT Madras and IIT Kanpur, India}}

\begin{abstract}
Grain boundary sliding models of superplastic flow take the elementary
event to be the shear of an oblate-spheroidal unit embedded in a solid
matrix, whose rate follows transition state theory. The mechanics of
this unit, and the assumptions linking it to dislocations and
macroscopic flow, have not been stated explicitly. Here the unit is
analysed as an Eshelby inclusion, a continuum, mathematical treatment
without using physical concepts. The analysis gives closed-form shear and
volumetric energy coefficients ($0.4456$ and $0.8889$ for aspect ratio
$1/2$ and Poisson ratio $1/3$) and shows that the shear stress of
the unit is a force dipole, decaying as $r^{-3}$ and changing sign with
direction, unlike a dislocation. The inputs,
assumptions and proposals needed to connect this unit of shear to
dislocation theory and superplastic flow are then stated and
classed as experimentally validated, physically realistic or fitting
constants. Under three such proposals, a row of about four units acquires
the Peierls--Nabarro profile of a full dislocation, and two to three
units that of a Shockley partial; the Peierls and Frank--Read stresses
and the stacking-fault energy then follow from experimental inputs. With the
unit energy as the free energy of activation and strains fitted to
superplastic data, the flow law reproduces all 52 measured strain rates
of a Cu alloy within an order of magnitude. Molecular-dynamics
calculations for Cu outline the extension to dislocation emission,
mesoscopic grain boundary sliding and the inverse Hall--Petch effect. This
gives phenomenological models of superplastic flow a mechanical basis.
\end{abstract}

\begin{keyword}
Oblate-spheroidal Eshelby inclusion \sep Shear transformation \sep
Grain boundary sliding \sep Peierls--Nabarro model \sep Superplasticity
\sep Inverse Hall--Petch effect
\end{keyword}
\end{frontmatter}

\section{Introduction}
\label{sec:intro}
Dislocation theory describes crystal plasticity through line defects,
their motion and interactions~\citep{taylor1934,hirth1982}. Localised
rearrangements also provide a useful description of deformation in
glasses and disordered grain-boundary regions~\citep{argon1979,pad-part1,
falk-langer1998,homer-schuh2009}. An elastic eigenstrain inclusion represents
the change of shape associated with such a rearrangement, while its
irreversibility and kinetics require additional physical information.

A Burgers vector, which measures the incompatibility of the elastic
distortion with a fixed circuit and sign convention, may be written
\begin{equation}
 b_i=\oint_C\beta^{\mathrm e}_{ij}\,\mathrm dx_j.
 \label{eq:burgers}
\end{equation}
The integral of the gradient of a globally single-valued total
displacement is zero. Perfect lattice dislocations have translation
vectors of the crystal, whereas partial dislocations can have fractional
Burgers vectors and bound faults. Continuum dislocation descriptions do
not require atom-by-atom lattice circuits, and grain boundaries can
support dislocations and disconnections. Crystalline ceramics and the
crystalline regions of semicrystalline polymers retain reference
lattices even where glide is difficult. Thus restricted glide and
absence of crystalline order are distinct physical circumstances.

Eshelby's inclusion theory~\citep{eshelby1957,mura1987} supplies an
explicit elastic model for a local transformation. We use an oblate
spheroid with specified radius, aspect ratio, and eigenstrain. These
are model inputs, together with the elastic constants; their values
are not selected by isotropic elasticity. In particular, elasticity
alone does not select a crystallographic slip system or a transition
state. Continuum accuracy at a radius comparable with an atomic spacing
must be assessed against atomistic calculations.

The present analysis builds on the oblate-spheroid model of grain
boundary sliding~\citep{padmanabhan1977,pad-part1}, in which the
elementary sliding event is an oblate-spheroidal unit whose rate is
given by transition state theory. In contrast, Eshelby's analysis is a mathematical
treatment of the elastic field of an inclusion of ellipsoidal shape, of
which the oblate spheroid is one among a few shapes considered; it
contains no
kinetics. The connection of the shear of an oblate-spheroidal unit with
transition state theory, and through it with the rate of boundary
sliding, is not part of Eshelby's analysis and was introduced by
\citet{padmanabhan1977} and \citet{pad-part1}. Transition state theory
is therefore taken as a part of our model. Our aim is to supply the elastic mechanics
of that unit and to state explicitly the additional proposals and
experimental inputs that connect it with dislocation theory and with the
mesoscopic boundary-sliding analyses of \citet{pad-part1},
\citet{sripathi2014} and \citet{harisankar2025}.
The paper is in two parts. Part~A considers only a single oblate
spheroid and the properties that follow from its elastic analysis:
Section~\ref{sec:ostz} derives its elastic energy,
Section~\ref{sec:dipole} its far-field stress, and
Section~\ref{sec:single} the properties of one unit that follow from
these results, including how it compares with a line defect. Part~B
connects these results with dislocation theory and with superplastic
flow. Section~\ref{sec:assumptions} first collects all the inputs,
assumptions and proposals that this requires.
Section~\ref{sec:emergence} derives the slip profile of a row of units,
Section~\ref{sec:derived} estimates quantities of dislocation theory,
Section~\ref{sec:gb} the boundary energy,
Section~\ref{sec:stat} the interaction between units and
Section~\ref{sec:sinh} the flow law at a single boundary.
Section~\ref{sec:validation} compares the results with experiment and with
fitted superplastic-flow parameters, and Section~\ref{sec:meso} sets out
the route from a single boundary to mesoscopic superplastic flow and the
inverse Hall--Petch effect, with molecular-dynamics calculations for Cu in
\ref{app:extension}.

\subsection{Relation to earlier descriptions}
\label{sec:novelty}
Shear Transformation Zone (STZ) descriptions already connect local
rearrangements with elastic interactions~\citep{padmanabhan1977,argon1979,pad-part1,homer-schuh2009}. Peierls--Nabarro models
combine continuum elasticity with a supplied misfit-energy surface to
calculate dislocation profiles~\citep{peierls1940,nabarro1947,lu2000gsf,
xiang2008}. The present inclusion calculation supplies the elastic part
of such descriptions for one explicit event geometry; the misfit surface
and the kinetic information enter as separate inputs.

\parthead{Part A. Analysis and properties of a single oblate-spheroidal unit}

Part~A treats one oblate-spheroidal unit in an infinite isotropic matrix.
It asks how much elastic energy the unit stores when it shears and
dilates, and what stress it sets up in its surroundings. Nothing in
Part~A depends on dislocation theory or on superplasticity; these
connections are made in Part~B.

\section{\rev{The oblate-spheroidal transformation zone (\OSTZ) model}}
\label{sec:ostz}

\begin{figure}[!htbp]
\centering
\includegraphics[width=\linewidth]{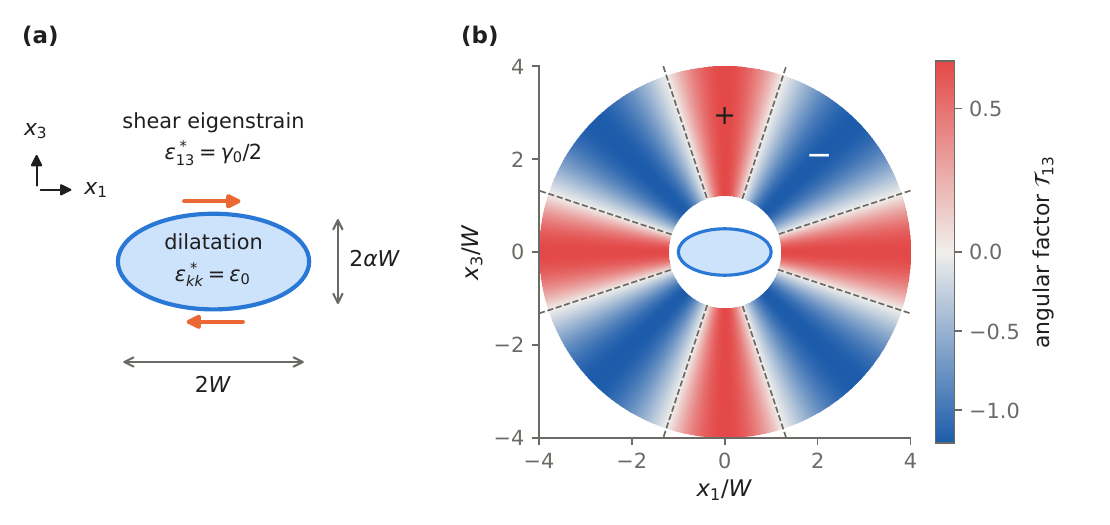}
\caption{(a) The oblate-spheroidal transformation zone: in-plane radius
$W$, half-thickness $\alpha W$, uniform shear eigenstrain
$\varepsilon^*_{13}=\gamma_0/2$ and dilatation $\varepsilon_0$, in an
infinite isotropic matrix. (b) Angular factor $\mathcal T_{13}$ of the
leading far-field shear stress $\sigma_{13}\propto\mathcal T_{13}/r^3$ in
the $x_1$--$x_3$ plane for $\nu=1/3$. Red and blue sectors have opposite
signs; dashed lines are nodal directions. The inner region, where the
point-dipole approximation does not apply, is left blank.}
\label{fig:schematic}
\end{figure}

This section defines the unit and derives its stored elastic energy,
which later sets the scale of the free energy of activation for the flow
law. An \OSTZ\ is an idealised local shear transformation, represented
by a uniform eigenstrain in an oblate spheroid; its size and
eigenstrains are inputs taken from the boundary-sliding model.
Figure~\ref{fig:schematic} shows the geometry, the prescribed
eigenstrains, and the sign pattern of the leading far-field shear stress
derived in Section~\ref{sec:dipole}.

\subsection{Geometric definition}
\label{sec:geometry}

We adopt an in-plane radius $W$ and take the aspect ratio
$\alpha=1/2$ as a modelling choice. Smoothness of the Eshelby solution
does not single out this aspect ratio; this is a choice of our model. The semi-axes and volume are

\begin{align}
  a_1=a_2=W,\quad a_3=W/2,&\quad \alpha\equiv a_3/a_1=\tfrac{1}{2},
  \notag\\
  V_0=\tfrac{4\pi}{3}a_1^2 a_3&=\tfrac{2\pi}{3}W^3.
  \label{eq:geometry}
\end{align}
The equatorial plane is the $x_1$--$x_2$ plane. The spheroid has a
curved surface, with maximum thickness $W$. We prescribe shear in
the $x_1$ direction across $x_3$. The
eigenstrain that produces this sliding, together with the
accompanying volume change measured by $\varepsilon_0$, is
\begin{equation}
  \varepsilon^*_{13}=\varepsilon^*_{31}=\gz/2\;(\text{shear}),\quad
  \varepsilon^*_{ii}=\varepsilon_0/3\;(\text{no sum; dilat.})
  \label{eq:eigenstrain}
\end{equation}

As the eigenstrain is a pure $13$-shear rather than an in-plane
($12$) shear, this geometry makes $S_{1313}$, not $S_{1212}$, the
Eshelby tensor component governing the stored elastic energy:
the central quantity computed in the rest of this section.

\subsection{Eshelby solution for the \OSTZ}
\label{sec:eshelby}

The inclusion and matrix are assumed to have the same isotropic
elastic constants, with $G>0$ and $-1<\nu<1/2$, in an infinite medium.
Eshelby's solution describes the response to a prescribed transformation;
it does not determine the atomic pathway or the free energy of activation.

We now need the actual stress and strain that the eigenstrain of
Equation~\eqref{eq:eigenstrain} produces inside the \OSTZ, since this is
what determines how much elastic energy is stored when the \OSTZ\
activates (Section~\ref{sec:energy}). General inclusion shapes require a more involved elasticity
boundary-value calculation; a useful feature of an \emph{ellipsoid} is that
Eshelby's inclusion theorem~\citep{eshelby1957} gives the answer in
closed form. The theorem states that the strain and stress inside a
homogeneous ellipsoidal inclusion are \emph{uniform} (independent of
position within the inclusion):
\begin{align}
  \varepsilon_{ij}^\mathrm{in}&=S_{ijkl}\varepsilon^*_{kl},
    \label{eq:eshelby1}\\
  \sigma_{ij}^\mathrm{in}&=C_{ijkl}(S_{klmn}-I_{klmn})\varepsilon^*_{mn}.
    \label{eq:eshelby2}
\end{align}
All of the shape-dependence is packaged into one object: the Eshelby
tensor is $S_{ijkl}$, which is constant inside an ellipsoidal inclusion,
so that $\varepsilon^\mathrm{in}_{ij}$ is uniform for the
uniform-eigenstrain problem considered here.

For the shear part of Equation~\eqref{eq:eigenstrain}, only $S_{1313}$ is needed; for the
general ellipsoid this component is built from two auxiliary shape
integrals, $I_\alpha$ and $I_{\alpha\beta}$, that depend only on the
semi-axes~\citep{mura1987}:
\begin{equation}
  S_{1313}=\frac{(a_1^2+a_3^2)I_{13}+(1-2\nu)(I_1+I_3)}{16\pi(1-\nu)},
  \label{eq:S1313}
\end{equation}
where $\Delta(s)=\sqrt{(a_1^2{+}s)(a_2^2{+}s)(a_3^2{+}s)}$,
\begin{align}
  I_\alpha&=2\pi a_1 a_2 a_3\int_0^\infty
    \frac{\mathrm ds}{(a_\alpha^2+s)\Delta(s)},
    \label{eq:Ialpha}\\
  I_{\alpha\beta}&=2\pi a_1 a_2 a_3\int_0^\infty
    \frac{\mathrm ds}{(a_\alpha^2+s)(a_\beta^2+s)\Delta(s)},\notag
\end{align}
with sum rule $I_1+I_2+I_3=4\pi$. These integrals are elementary but
tedious; the next three subsections evaluate them explicitly for our
oblate spheroid ($a_1=a_2=1$, $a_3=\alpha$ in reduced units), first
deriving $I_1$ in closed form, then $I_{13}$, then substituting
numbers for $\alpha=1/2$ to obtain the constraint factor $\beta_1$
that enters the elastic energy of Equation~\eqref{eq:dF0}.

\subsubsection{\texorpdfstring{Derivation of $I_1$}{Derivation of I1}}
\label{sec:I1}

The integral $I_1$ measures how strongly the matrix constrains the unit
in its own plane. A closed form is derived so that $S_{1313}$, and hence
the shear energy, can be evaluated for any aspect ratio rather than for
one value only.

\paragraph{Starting point}
Focus on $a_1=a_2=1$, $a_3=\alpha$. Then
$\Delta(s)=(1+s)\sqrt{\alpha^2+s}$ and
\begin{equation}
  I_1=2\pi\alpha\int_0^\infty\frac{\mathrm ds}
  {(1+s)^2\sqrt{\alpha^2+s}}.
  \label{eq:I1start}
\end{equation}

\paragraph{Substitution}
Let $t=\sqrt{\alpha^2+s}$, so $s=t^2-\alpha^2$, $\mathrm ds=2t\,\mathrm dt$,
$\sqrt{\alpha^2+s}=t$, and $1+s=t^2+(1-\alpha^2)\equiv t^2+e^2$
(where $e^2\equiv1-\alpha^2$). New limits: $t:\alpha\to\infty$.
Substituting:
\begin{equation}
  I_1=4\pi\alpha\int_\alpha^\infty\frac{\mathrm dt}{(t^2+e^2)^2}.
  \label{eq:I1sub}
\end{equation}

\paragraph{Antiderivative}
The reduction formula (verified by differentiating the right side):
\begin{equation}
  \int\frac{\mathrm dt}{(t^2+e^2)^2}=
  \frac{t}{2e^2(t^2+e^2)}+\frac{\arctan(t/e)}{2e^3}+C.
  \label{eq:antideriv}
\end{equation}
\paragraph{Derivation} Let $F=t/[2e^2(t^2{+}e^2)]$,
$G_f=\arctan(t/e)/(2e^3)$:
\begin{align*}
  F'&=\frac{1}{2e^2}\cdot\frac{(t^2+e^2)-t\cdot2t}{(t^2+e^2)^2}
    =\frac{e^2-t^2}{2e^2(t^2+e^2)^2},\\
  G_f'&=\frac{1}{2e^3}\cdot\frac{1/e}{1+(t/e)^2}
    =\frac{1}{2e^2(t^2+e^2)}.
\end{align*}
Sum: $F'+G_f'=[(e^2-t^2)+(t^2+e^2)]/[2e^2(t^2+e^2)^2]
=1/(t^2+e^2)^2$. $\square$

\paragraph{Evaluating at the bounds}
\textit{Upper bound} ($t\to\infty$): $F\to0$, $\arctan\to\pi/2$, so
$[F+G_f]_{t\to\infty}=\pi/(4e^3)$.

\textit{Lower bound} ($t=\alpha$): note $\alpha^2+e^2=1$, so
$F|_\alpha=\alpha/(2e^2)$,
$G_f|_\alpha=\arctan(\alpha/e)/(2e^3)$. Hence:
\begin{align}
  \int_\alpha^\infty\frac{\mathrm dt}{(t^2+e^2)^2}
  &=\frac{\pi}{4e^3}-\frac{\alpha}{2e^2}
    -\frac{\arctan(\alpha/e)}{2e^3}\notag\\
  &=\frac{1}{2e^3}\!\left[\frac{\pi}{2}-\arctan\!\left(\frac{\alpha}{e}
    \right)\right]-\frac{\alpha}{2e^2}.
  \label{eq:I1bounds}
\end{align}

\paragraph{Identity $\pi/2-\arctan(\alpha/e)=\arccos\alpha$}
Set $\alpha=\cos\phi$ ($e=\sin\phi$). Then
$\arctan(\cos\phi/\sin\phi)=\arctan(\cot\phi)=\pi/2-\phi$, so
$\pi/2-\arctan(\alpha/e)=\phi=\arccos\alpha$. Substituting into
Equation~\eqref{eq:I1bounds} and multiplying by $4\pi\alpha$, and
replacing $e^3=(1-\alpha^2)^{3/2}$:
\begin{equation}
  \boxed{I_1=\frac{2\pi\alpha}{(1-\alpha^2)^{3/2}}
  \!\left[\arccos\alpha-\alpha\sqrt{1-\alpha^2}\right],\qquad
  I_3=4\pi-2I_1.}
  \label{eq:I1result}
\end{equation}

\subsubsection{\texorpdfstring{Derivation of $I_{13}=(I_3-I_1)/(1-\alpha^2)$}{Derivation of I13 from I3 and I1}}
\label{sec:I13}

Equation~\eqref{eq:S1313} for $S_{1313}$ still needs the
cross-integral $I_{13}$, which superficially requires evaluating yet
another integral of the same type as $I_1$. In fact $I_{13}$ can be
obtained algebraically from $I_1$ and $I_3$ without any new
integration, by exploiting the way the two denominators
$(1+s)$ and $(\alpha^2+s)$ combine. Writing the integrals explicitly,
with $\Delta=(1+s)(\alpha^2+s)^{1/2}$,
\begin{align}
  I_1&=2\pi\alpha\int_0^\infty\frac{\mathrm ds}
    {(1+s)^2\sqrt{\alpha^2+s}},\notag\\
  I_3&=2\pi\alpha\int_0^\infty\frac{\mathrm ds}
    {(1+s)(\alpha^2+s)^{3/2}},\notag\\
  I_{13}&=2\pi\alpha\int_0^\infty\frac{\mathrm ds}
    {(1+s)^2(\alpha^2+s)^{3/2}}.\notag
\end{align}
Forming $I_3-I_1$ and factoring out $1/[(1+s)\sqrt{\alpha^2+s}]$:
\begin{align*}
  I_3-I_1&=2\pi\alpha\int_0^\infty\frac{1}{(1+s)\sqrt{\alpha^2+s}}
  \!\left[\frac{1}{\alpha^2+s}-\frac{1}{1+s}\right]\mathrm ds.
\end{align*}
The bracket equals $(1-\alpha^2)/[(\alpha^2{+}s)(1{+}s)]$:
\begin{align*}
  I_3-I_1&=(1-\alpha^2)\cdot2\pi\alpha\int_0^\infty
  \frac{\mathrm ds}{(1+s)^2(\alpha^2+s)^{3/2}}\\
  &=(1-\alpha^2)I_{13}.
\end{align*}
Therefore:
\begin{equation}
  \boxed{I_{13}=\frac{I_3-I_1}{1-\alpha^2}.}\quad\square
  \label{eq:I13result}
\end{equation}

\subsubsection{\texorpdfstring{Numerical evaluation ($\alpha=1/2$, $\nu=1/3$)}{Numerical evaluation (alpha=1/2, nu=1/3)}}
\label{sec:numeval}

With $I_1$ and $I_{13}$ available in closed form, we insert the adopted
aspect ratio $\alpha=1/2$ and a representative Poisson ratio $\nu=1/3$
to obtain $\beta_1$, the coefficient of the shear part of the elastic
energy (Equation~\eqref{eq:Eshear} below).

\begin{figure}[!htbp]
\centering
\includegraphics[width=\linewidth]{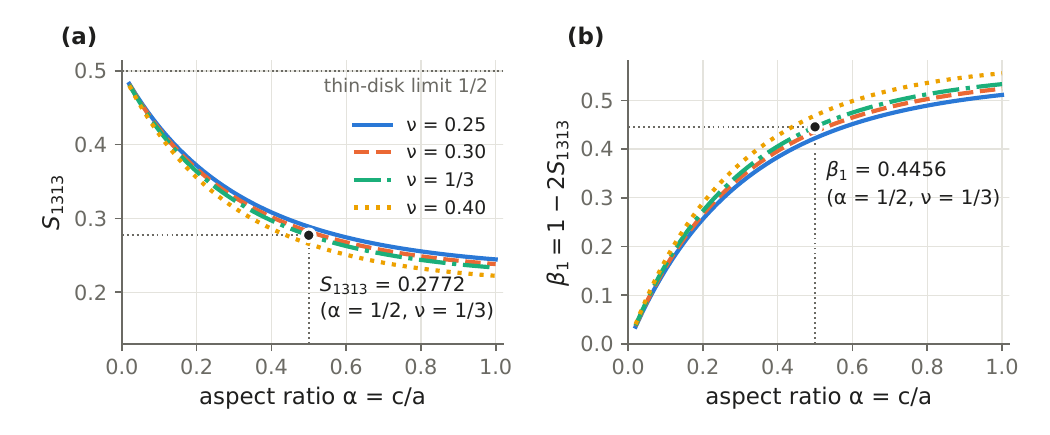}
\caption{(a) Eshelby shear component $S_{1313}$ and (b) shear energy
coefficient $\beta_1=1-2S_{1313}$ of a homogeneous oblate spheroid versus
aspect ratio $\alpha=c/a$ for four Poisson ratios. The curves approach the
fixed-eigenstrain thin-disk limit $S_{1313}=1/2$ and the sphere value
$(4-5\nu)/[15(1-\nu)]$. The filled circle, with dotted guide lines to
both axes, marks the values used in this work ($\alpha=1/2$, $\nu=1/3$):
$S_{1313}=0.2772$ and $\beta_1=0.4456$.}
\label{fig:s1313}
\end{figure}

Direct evaluation, retaining precision until the final step, gives
\begin{align*}
 I_1&=2.97068824,& I_3&=6.62499414,& W^2 I_{13}&=4.87240787,\\
 S_{1313}&=0.27720028,& \beta_1=1-2S_{1313}&=0.44559943.
\end{align*}
The reduced-coordinate derivations above set $W=1$; in physical units
$I_{13}$ has dimensions of inverse length squared.
Figure~\ref{fig:s1313} shows how $S_{1313}$ and $\beta_1$ vary with
aspect ratio and Poisson ratio. Poisson's ratio is a material property,
and the aspect ratio of the unit is expected to depend on the
misorientation of the boundary. $\beta_1$ can therefore be regarded as a
property of the material and of the boundary type, although not as a
universal constant. The same coefficient was used by \citet{pad-part1}
for the same aspect ratio: their $\beta_1=0.470(1.590-\nu)/(1-\nu)$, with
$\beta_2$ as in Equation~\eqref{eq:beta2}, has the Poisson-ratio
dependence of Table~\ref{tab:eshelby} and equals $1.99\beta_1$ for every
$\nu$ listed there ($0.886$ at $\nu=1/3$, against $2\beta_1=0.891$). The
later coefficient $0.944(1.590-\nu)/(1-\nu)$ of \citet{harisankar2025},
written for the tensor shear strain of the von Mises treatment, equals
$3.99\beta_1$, i.e.\ $4\beta_1$ to within $0.2\%$ (Supporting Information,
Section~S6). The present value is therefore equivalent to the earlier
ones once the strain measure of each paper is taken into account.

\subsubsection{Limiting-case verification}
\label{sec:limits}

Before trusting $\beta_1=0.446$ as a physically reasonable
constraint factor, it is worth checking the formula against two
shapes whose answer we already know by symmetry or by elementary
physical reasoning: the sphere, where isotropy gives an independent
check, and the vanishingly thin disk, for fixed eigenstrain and radius.

\paragraph{Sphere ($\alpha\to1$)}
For a sphere all semi-axes equal: $I_1=I_2=I_3=4\pi/3$.
Direct integration:
$I_{13}^\text{sphere}=2\pi\int_0^\infty(1+s)^{-7/2}\mathrm ds
=2\pi[u^{-5/2}/(-5/2)]_1^\infty=4\pi/5$.
Substituting ($1+\alpha^2\to2$, $I_1+I_3=8\pi/3$):
\begin{equation}
  S_{1313}^\text{sphere}=\frac{4-5\nu}{15(1-\nu)}.
  \label{eq:S1313sphere}
\end{equation}
For $\nu=1/3$: $S_{1313}^\text{sphere}=7/30=0.233$,
$\beta_1^\text{sphere}=8/15\approx0.533$.

\paragraph{Thin disk ($\alpha\to0$)}
$I_3\to4\pi$, $I_{13}\to4\pi$, $S_{1313}^\text{disk}=\half$,
$\beta_1\to0$. This is a vanishing-thickness inclusion at fixed
eigenstrain; it is not a traction-free crack at fixed displacement jump.
The OSTZ aspect ratio $\alpha=1/2$ gives an intermediate value.
The Supporting Information reports independent numerical checks.

\subsubsection{\texorpdfstring{Dilatational constraint factor $\beta_2$}{Dilatational constraint factor beta2}}
\label{sec:beta2}

The unit also expands slightly, in a transient manner, while it shears
and reaches an adjacent equivalent minimum energy configuration. The coefficient $\beta_2$
measures the elastic energy stored because the matrix resists this
volume change; it is the second term of the activation energy.

For pure volumetric eigenstrain, the total elastic self-energy of a
homogeneous inclusion in an infinite isotropic medium is independent of
ellipsoidal shape. This energy invariance also follows from the homogeneous case in
\citet{bohm2022}. The interior strain need not be isotropic for an
oblate spheroid; its trace is shape independent. The sphere therefore
provides a convenient way to evaluate the energy contraction, without
assuming that the oblate interior stress is hydrostatic. For a sphere,

\begin{equation}
  S_{ijkk}^\text{sphere}=\frac{1+\nu}{3(1-\nu)}\delta_{ij}.
  \label{eq:Sijkk}
\end{equation}
This follows from the known sphere values~\citep{mura1987}
$S_{1111}=(7-5\nu)/[15(1-\nu)]$,
$S_{1122}=(5\nu-1)/[15(1-\nu)]$, summing:
$S_{1111}+S_{1122}+S_{1133}=(1+\nu)/[3(1-\nu)]$. Full derivation
of $\beta_2$ follows in Section~\ref{sec:Edilat}; the result:
\begin{equation}
  \boxed{\beta_2=\frac{4(1+\nu)}{9(1-\nu)}.}
  \label{eq:beta2}
\end{equation}
For $\nu=1/3$: $\beta_2=4(4/3)/[9(2/3)]=0.889$.
Like $\beta_1$, $\beta_2$ is a material-dependent constant, since it
is set by the Poisson ratio of the material. In the boundary-sliding
model the dilatation is also tied to the shear strain through the von
Mises relation $\varepsilon_0=\gz/\sqrt3$, with $\gz$ a tensor shear
strain~\citep{harisankar2025}, so that the dilatational term of the
energy is fixed once the shear term is fixed; in that sense $\beta_2$ is a
second material-dependent constant linked to $\beta_1$.

The expression involving $S_{3333}$ alone is not the energy coefficient
for the isotropic volumetric eigenstrain in Equation~\eqref{eq:eigenstrain}.
The required contraction is $S_{iikk}$, not one longitudinal component.
Table~\ref{tab:eshelby} collects the Eshelby components for $\alpha=1/2$.

\begin{table}[!htbp]
\centering
\small
\caption{Eshelby components for $\alpha=1/2$.}
\label{tab:eshelby}
\begin{tabularx}{\textwidth}{CCCCC}
\toprule
$\nu$ & $S_{1313}$ & $S_{1212}$ & $S_{3333}$ & $\beta_1$\\
\midrule
0.30 & 0.2822 & 0.1769 & 0.7264 & 0.4356\\
1/3  & 0.2772 & 0.1739 & 0.7364 & 0.4456\\
0.35 & 0.2745 & 0.1723 & 0.7418 & 0.4510\\
0.40 & 0.2656 & 0.1670 & 0.7596 & 0.4688\\
\bottomrule
\end{tabularx}
\end{table}

\subsection{Constraint-factor conventions}
This subsection states how $\beta_1$ is to be used when the present
energy is compared with other treatments of the same unit. For the
specified homogeneous inclusion, $\beta_1=1-2S_{1313}$ applies
in both crystal and boundary calculations. An effective value near unity
is a separate constitutive assumption, not a consequence of replacing
$G$ by a softened grain-boundary modulus. If the reference modulus is
changed while preserving an energy, the product $\beta_1G\gamma_0^2$
must be preserved. A real inclusion--matrix modulus contrast requires
an inhomogeneity calculation. Engineering and tensor shear conventions
also differ by a factor of two in strain and four in the quadratic
coefficient; they must be converted consistently in both energy and work.

\subsection{OSTZ elastic strain energy}
\label{sec:energy}

The elastic self-energy of the inclusion and its surrounding matrix is
$E_{\rm el}=-\half\sigma_{ij}^{\rm in}\varepsilon^*_{ij}V_0$.
The shear--dilatation cross term vanishes for the specified shear aligned
with the spheroid axes by reflection symmetry. Isotropy of the stiffness
alone would not justify this statement for an arbitrary inclusion and
arbitrary deviatoric eigenstrain. For $\gz=0.10$, $\varepsilon_0=0.05$,
and $\nu=1/3$, the volumetric term is approximately one third of the total.
The notation $\dF$ in the elastic derivations below denotes this elastic
energy scale; the kinetic barrier $F^\ddagger$ will be distinguished explicitly.
In the model of \citet{pad-part1}, the volume expansion accompanying an
elementary shear transformation is transient and vanishes on reaching
the next local minimum energy configuration, whereas some induced shear can remain
until its surroundings also relax. Thus $\varepsilon_0$ can represent a
trial activation-path eigenstrain without predicting a persistent
post-event volume increase. The elastic calculation does not itself
determine whether such a transient dilation occurs; it is assumed to
occur because the shear takes place in a unit embedded in a solid
matrix.

\subsubsection{Part~1: Shear contribution}
\label{sec:Eshear}

The matrix prevents the unit from shearing freely. The four steps below
compute the stress that this constraint produces inside the unit and the
elastic energy it stores, which is the shear part of the activation
energy.

Eigenstrain $\varepsilon^*_{13}=\varepsilon^*_{31}=\gz/2$, all other
components are zero.

\paragraph{Step~1 (constrained strain)}
Equation~\eqref{eq:eshelby1} gives $\varepsilon_{13}^\mathrm{in}$ as a
contraction of the Eshelby tensor with the \emph{full} eigenstrain
tensor, so both nonzero components $\varepsilon^*_{13}$ and
$\varepsilon^*_{31}$ contribute:
$\varepsilon_{13}^\mathrm{in}=S_{1313}\varepsilon^*_{13}+S_{1331}\varepsilon^*_{31}$.
By the minor symmetry of the Eshelby tensor, $S_{1313}=S_{1331}$ (a
direct consequence of the symmetry $\varepsilon^*_{kl}=\varepsilon^*_{lk}$
of any physical strain tensor); the two terms combine:
$\varepsilon_{13}^\mathrm{in}=S_{1313}(\gz/2+\gz/2)=S_{1313}\gz$.

\paragraph{Step~2 (elastic strain)}
The eigenstrain $\varepsilon^*_{13}$ is, by definition, a stress-free
transformation strain; it is the additional strain
$e_{13}^\mathrm{el}=\varepsilon_{13}^\mathrm{in}-\varepsilon^*_{13}$,
the difference between the actual (constrained) strain and the
eigenstrain, that the surrounding matrix elastically resists and
that therefore generates stress via Hooke's law:
$e_{13}^\mathrm{el}=\varepsilon_{13}^\mathrm{in}-\varepsilon^*_{13}
=(S_{1313}-\half)\gz$.

\paragraph{Step~3 (interior stress)}
For an isotropic matrix the relevant stiffness component is simply
$C_{1313}=C_{1331}=G$ (the standard
isotropic-elasticity relation between shear stress and shear strain),
so
$\sigma_{13}^\mathrm{in}=2Ge_{13}^\mathrm{el}
=2G(S_{1313}-\half)\gz$.

\paragraph{Step~4 (energy contraction)}
Both $(1,3)$ and $(3,1)$ contribute equally:
\[
  \dF^\text{shear}=-\half(\sigma_{13}^\mathrm{in}\varepsilon^*_{13}
  +\sigma_{31}^\mathrm{in}\varepsilon^*_{31})V_0
  =-\half\sigma_{13}^\mathrm{in}\gz V_0.
\]
Substituting:
$\dF^\text{shear}=-\half\cdot2G(S_{1313}-\half)\gz\cdot\gz V_0
=G(\half-S_{1313})\gz^2 V_0$.
\begin{equation}
  \boxed{\dF^\text{shear}=\half\beta_1 G\gz^2 V_0,\quad
  \beta_1\equiv1-2S_{1313}.}
  \label{eq:Eshear}
\end{equation}
Since $S_{1313}<\half$ for any physical oblate spheroid,
$\beta_1>0$ and the shear energy is positive definite.

\subsubsection{Part~2: Dilatational contribution}
\label{sec:Edilat}

The same four steps give the energy stored because the matrix resists the
transient volume increase of the unit.

The same energy bookkeeping is now repeated for the volumetric part
of the eigenstrain, using the spherical-inclusion Eshelby result
$S_{ijkk}^\text{sphere}$ already obtained in
Section~\ref{sec:beta2} (Equation~\eqref{eq:Sijkk}).
Eigenstrain $\varepsilon^*_{ij}=(\varepsilon_0/3)\delta_{ij}$.

\paragraph{Step~1 (constrained strain inside a sphere)}
From Equations~\eqref{eq:eshelby1} and \eqref{eq:Sijkk}:
\begin{equation}
  \varepsilon_{ij}^\mathrm{in}=S_{ijkl}\varepsilon^*_{kl}
  =\frac{\varepsilon_0}{3}\cdot\frac{1+\nu}{3(1-\nu)}\delta_{ij}
  =\frac{(1+\nu)\varepsilon_0}{9(1-\nu)}\delta_{ij}.
  \label{eq:constrained_dil}
\end{equation}

\paragraph{Step~2 (elastic strain)}
\begin{align}
  e_{ij}^\mathrm{el}&=\varepsilon_{ij}^\mathrm{in}-\varepsilon^*_{ij}
    =\varepsilon_0\!\left[\frac{1+\nu}{9(1-\nu)}-\frac{1}{3}\right]\delta_{ij}
    \notag\\
  &=\varepsilon_0\cdot\frac{(1+\nu)-3(1-\nu)}{9(1-\nu)}\delta_{ij}
    =-\frac{2(1-2\nu)\varepsilon_0}{9(1-\nu)}\delta_{ij}.
    \label{eq:edilelastic}
\end{align}
(Numerator: $1+\nu-3+3\nu=4\nu-2=-2(1-2\nu)$.)
Volumetric trace:
$e_{kk}^\mathrm{el}=-6(1-2\nu)\varepsilon_0/[9(1-\nu)]
=-2(1-2\nu)\varepsilon_0/[3(1-\nu)]$.

\paragraph{Step~3 (interior stress)}
Hooke's law is $\sigma_{ij}=\lambda e_{kk}\delta_{ij}+2Ge_{ij}$.
The second term is not purely deviatoric; for this spherical dilatation
both terms are hydrostatic.
Using $\lambda=2G\nu/(1-2\nu)$:
\begin{align}
  \lambda e_{kk}^\mathrm{el}
    &=\frac{2G\nu}{1-2\nu}\cdot\frac{-2(1-2\nu)\varepsilon_0}{3(1-\nu)}
    =-\frac{4G\nu\varepsilon_0}{3(1-\nu)},\notag\\
  2Ge_{11}^\mathrm{el}
    &=-\frac{4G(1-2\nu)\varepsilon_0}{9(1-\nu)}.
    \notag
\end{align}
Adding the two Lam\'e contributions, with common denominator
$9(1-\nu)$:
\begin{align}
  \sigma_{11}^\mathrm{in}&=-\frac{12G\nu\varepsilon_0}{9(1-\nu)}
    -\frac{4G(1-2\nu)\varepsilon_0}{9(1-\nu)}\notag\\
  &=-\frac{G\varepsilon_0}{9(1-\nu)}[12\nu+4-8\nu]
    =-\frac{4G(1+\nu)\varepsilon_0}{9(1-\nu)}.
    \label{eq:sig11in}
\end{align}

\paragraph{Step~4 (energy contraction)}
By isotropy all three diagonal stress components are equal,
$\sigma_{11}^\mathrm{in}=\sigma_{22}^\mathrm{in}=\sigma_{33}^\mathrm{in}$,
so the double contraction with the diagonal eigenstrain
$\varepsilon^*_{ii}=\varepsilon_0/3$ collapses to three identical
terms: $\sigma_{ij}^\mathrm{in}\varepsilon^*_{ij}
=3\sigma_{11}^\mathrm{in}(\varepsilon_0/3)=\sigma_{11}^\mathrm{in}\varepsilon_0$.
Substituting into the energy theorem,
\begin{equation}
  \boxed{\dF^\text{dilat}=\half\beta_2 G\varepsilon_0^2 V_0,\quad
  \beta_2=\frac{4(1+\nu)}{9(1-\nu)}.}
  \label{eq:Edilat}
\end{equation}
For stable isotropic elasticity, $\beta_2>0$; it exceeds unity when
$\nu>5/13$ and tends to $4/3$ as $\nu\to1/2$. It is normalised by $G$,
not by the bulk modulus, so it is not a fraction bounded by one.

\subsubsection{Total elastic energy}

Adding the two contributions, using the reflection symmetry noted above,

\begin{equation}
  \boxed{\dF=\half(\beta_1\gz^2+\beta_2\varepsilon_0^2)GV_0,\quad
  V_0=\tfrac{2\pi}{3}W^3.}
  \label{eq:dF0}
\end{equation}
Here $E_{\rm el}$ is the total elastic self-energy. An unconstrained
stress-free transformation stores zero elastic energy. The present
positive energy is lower than the energy with the transformation fully
clamped, but is not itself a transition-state calculation. The assumption
$F^\ddagger=E_{\rm el}$ can be used as an explicit kinetic approximation;
chemical, structural, and entropic contributions are not thereby derived.

\paragraph{Numerical check}
The purpose of this calculation is to give the size of the elastic energy
of one unit for illustrative inputs typical of a high-angle boundary in
Cu, to compare it with an independent atomistic energy scale, and to fix
the inputs used in the numerical examples of Part~B.
For $\gz=0.10$, $\varepsilon_0=0.05$, $W=0.75$~nm and $G=22$~GPa,
$V_0=0.883573$~nm$^3$ and $E_{\rm el}=0.40512$~eV.
This exceeds the illustrative $0.38$~eV reference by $6.61\%$.
The agreement of energy scales does not identify an activation pathway.
The free-volume increment is $\varepsilon_0V_0=0.04418$~nm$^3$ per
zone; per atom it is $\varepsilon_0\Omega$ if the zone contains
$V_0/\Omega$ atoms, or $0.05\Omega$ with these inputs.
Table~\ref{tab:canonical} lists these illustrative inputs and the derived
scales.

\begin{table}[!htbp]
\centering
\small
\caption{Illustrative \OSTZ\ inputs and derived scales; $\nu=1/3$
for the listed boundary coefficients. These values are not universal.}
\label{tab:canonical}
\begin{tabularx}{\textwidth}{>{\raggedright\arraybackslash}p{4cm}CCC}
\toprule
Parameter & Symbol & GB/superpl. & Crystal\\
\midrule
OSTZ radius & $W$ & $2.5b$ & $\approx b$\\
Shear eigenstrain & $\gz$ & 0.10 & $s_m$ (0.10--0.20)\\
Dilat.\ eigenstrain & $\varepsilon_0$ & 0.05 & ---\\
Shear constraint & $\beta_1$ & 0.446 & $1-2S_{1313}$\\
Dilat.\ constraint & $\beta_2$ & 0.889 & ---\\
Elastic energy (example) & $E_{\rm el}$ & 0.40512~eV & input dependent\\
Slip-amplitude ratio & $N_c$ & 4 & $b/(\gz W)$\\
\bottomrule
\end{tabularx}
\end{table}

\section{Far-Field Stress: The Dipole Structure}
\label{sec:dipole}

This section derives how the stress of one unit falls off with distance
and direction. The result, a force-dipole field valid for $r\gg W$,
controls how neighbouring units interact (Section~\ref{sec:stat}) and
shows how a row of units relates to a line defect
(Section~\ref{sec:nomacroslip}).

\subsection{Why a dipole, not a monopole}

The order of the leading term decides how quickly the influence of one
unit dies away with distance. The force monopole vanishes:
\begin{equation}
  F_i=\oint_{\partial V_0}\sigma_{ij}^* n_j\,\mathrm dS
  =C_{ijkl}\varepsilon^*_{kl}
  \underbrace{\oint_{\partial V_0}n_j\,\mathrm dS}_{=0}=0.
  \label{eq:monopole}
\end{equation}
(The surface integral of the outward normal over any closed surface
vanishes by the divergence theorem.) The leading far-field term is
therefore the \emph{force-dipole} at order $r^{-3}$.

The leading terms of the stress multipole expansion~\citep{mura1987} are
\begin{equation}
  \sigma_{ij}(\mathbf r)=
  \underbrace{F_k\mathcal G_{ijk}}_{\text{monopole }r^{-2}}
  -\underbrace{M_{kl}\partial_l\mathcal G_{ijk}}_{\text{dipole }r^{-3}}
  +\underbrace{Q_{klm}\partial_l\partial_m\mathcal G_{ijk}}_{r^{-4}}
  +\cdots
  \label{eq:multipole}
\end{equation}
Here $Q_{klm}$ includes the Taylor factor $1/2$ in the second force
moment. For the centred, uniform, inversion-symmetric inclusion it
vanishes; the next stress correction is of order $r^{-5}$. Since
$F_k=0$, the leading nonzero term is the dipole. The moment
tensor~\citep{mura1987} is
\begin{equation}
  M_{kl}=C_{ijkl}\varepsilon^*_{ij}V_0\;\Rightarrow\;
  M_{13}=M_{31}=G\gz V_0,
  \label{eq:moment}
\end{equation}
representing a shear double couple. The volumetric eigenstrain also
contributes $M_{ii}=K\varepsilon_0V_0$ (no sum), where
$K=2G(1+\nu)/[3(1-2\nu)]$. The following derivation isolates the shear
contribution. Here $\mathcal G_{ijk}$ in the stress expansion denotes
the stress response to a unit point force, distinct from the displacement
Green tensor $\mathcal G_{ij}$ below.

\subsection{Step A: Kelvin Green's function and its derivative}
\label{sec:stepA}

Far from the unit, its action on the matrix is that of a pair of opposed
force couples. Their field is built from the displacement produced by a
single point force, the Kelvin solution, and from its derivative, which
this step supplies.

The elastic displacement due to a unit point force at the origin in an
infinite isotropic medium (Kelvin--Somigliana solution~\citep{mura1987}):
\begin{equation}
  \mathcal G_{ij}(\mathbf r)=A\!\left[\frac{(3-4\nu)\delta_{ij}}{r}
  +\frac{x_ix_j}{r^3}\right],\quad A=\frac{1}{16\pi G(1-\nu)}.
  \label{eq:Gij}
\end{equation}

\paragraph{Differentiating the first term}
Since $\partial_l(1/r)=-x_l/r^3$:
\[
  \partial_l\!\left[\frac{(3-4\nu)\delta_{ij}}{r}\right]
  =-\frac{(3-4\nu)\delta_{ij}x_l}{r^3}.
\]

\paragraph{Differentiating the second term (product rule)}
$\partial x_i/\partial x_l=\delta_{il}$, $\partial(r^{-3})/\partial x_l=-3x_l/r^5$:
\[
  \partial_l\!\left(\frac{x_ix_j}{r^3}\right)
  =\frac{\delta_{il}x_j+\delta_{jl}x_i}{r^3}-\frac{3x_lx_ix_j}{r^5}.
\]

\paragraph{Combined result}
\begin{align}
  \frac{\partial\mathcal G_{ij}}{\partial x_l}
  =A\Bigl[&-\frac{(3-4\nu)\delta_{ij}x_l}{r^3}
    +\frac{\delta_{il}x_j+\delta_{jl}x_i}{r^3}
    \notag\\
  &-\frac{3x_lx_ix_j}{r^5}\Bigr].
  \label{eq:dG}
\end{align}

\subsection{Step B: Displacement from the moment tensor}
\label{sec:stepB}

This step gives the displacement of the matrix produced by the force
couples of the sheared unit, from which the stress follows in Step~C.

For an Eshelby eigenstrain inclusion, the far-field exterior
displacement is~\citep{eshelby1957,mura1987} $u_i=-M_{kl}\partial_l\mathcal G_{ik}$.
With $M_{13}=M_{31}=G\gz V_0$ (all other components are zero),
the only nonzero contributions are $(k,l)=(1,3)$ and $(3,1)$.

Substituting Equation~\eqref{eq:dG} (with $\delta_{13}=0$):
\begin{align}
  \frac{\partial\mathcal G_{i1}}{\partial x_3}
  &=A\!\left[-\frac{(3-4\nu)\delta_{i1}x_3}{r^3}
    +\frac{\delta_{i3}x_1}{r^3}-\frac{3x_3x_ix_1}{r^5}\right],
    \notag\\
  \frac{\partial\mathcal G_{i3}}{\partial x_1}
  &=A\!\left[-\frac{(3-4\nu)\delta_{i3}x_1}{r^3}
    +\frac{\delta_{i1}x_3}{r^3}-\frac{3x_1x_ix_3}{r^5}\right].
    \notag
\end{align}

Adding term by term, the $\delta_{i1}x_3$ coefficient is
$1-(3-4\nu)=4\nu-2=2(2\nu-1)$ (same for $\delta_{i3}x_1$),
and the $r^{-5}$ terms add to give $-6Ax_1x_3x_i/r^5$:
\begin{equation}
  u_i=-G\gz V_0 A\Bigl[\frac{2(2\nu-1)(\delta_{i1}x_3+\delta_{i3}x_1)}{r^3}
  -\frac{6x_1x_3x_i}{r^5}\Bigr].
  \label{eq:ui}
\end{equation}

Components relevant to $\sigma_{13}$:
\begin{align}
  u_1&=-G\gz V_0 A\!\left[\frac{2(2\nu-1)x_3}{r^3}
    -\frac{6x_1^2x_3}{r^5}\right],
    \label{eq:u1}\\
  u_3&=-G\gz V_0 A\!\left[\frac{2(2\nu-1)x_1}{r^3}
    -\frac{6x_1x_3^2}{r^5}\right].
    \label{eq:u3}
\end{align}

\subsection{Step C: Stress from displacement gradients}
\label{sec:stepC}

Step B gave the displacement field $u_i(\mathbf r)$ produced by the
\OSTZ\ dipole; what we actually need for the far-field stress
(Equation~\eqref{eq:dipole} below) is the shear stress $\sigma_{13}$, which
by Hooke's law is built from the symmetric displacement gradient
$\partial_3u_1+\partial_1u_3$. Because the shear component has
$\delta_{13}=0$, the $\lambda$-term (the dilatational part of Hooke's
law) drops out automatically, leaving a pure shear-modulus relation:
$\sigma_{13}=G(\partial_3u_1+\partial_1u_3)$. The remainder of this
subsection differentiates the two terms of $u_1$ and $u_3$ found in
Step B, term by term, and adds the results.

\paragraph{Auxiliary results}
$\partial_3(x_3/r^3)=(r^2-3x_3^2)/r^5$;
$\partial_3(x_3/r^5)=(r^2-5x_3^2)/r^7$; likewise for $x_1$.

\paragraph{Computing $\partial_3u_1$}
Differentiating Equation~\eqref{eq:u1} with respect to $x_3$:
\begin{align}
  \partial_3u_1&=-G\gz V_0 A\Bigl[\frac{2(2\nu-1)(r^2-3x_3^2)}{r^5}
    \notag\\
  &\phantom{=-G\gz V_0 A\Bigl[}\;-\frac{6x_1^2(r^2-5x_3^2)}{r^7}\Bigr].
    \label{eq:du1dx3}
\end{align}

\paragraph{Computing $\partial_1u_3$}
By the same procedure (swap $x_1\leftrightarrow x_3$):
\begin{align}
  \partial_1u_3&=-G\gz V_0 A\Bigl[\frac{2(2\nu-1)(r^2-3x_1^2)}{r^5}
    \notag\\
  &\phantom{=-G\gz V_0 A\Bigl[}\;-\frac{6x_3^2(r^2-5x_1^2)}{r^7}\Bigr].
    \label{eq:du3dx1}
\end{align}

\paragraph{Summing and converting to unit vectors $\hat r_i=x_i/r$}
$1/r^5$ group:
$2(2\nu-1)[2-3(\hat r_1^2+\hat r_3^2)]/r^3$.
$1/r^7$ group, expanding $(x_1^2+x_3^2)r^2-10x_1^2x_3^2$:
$[-6(\hat r_1^2+\hat r_3^2)+60\hat r_1^2\hat r_3^2]/r^3$.
Collecting the $(\hat r_1^2+\hat r_3^2)$ terms:
$-6(2\nu-1)-6=-12\nu$. Hence:
\begin{equation}
  \partial_3u_1+\partial_1u_3=
  -G\gz V_0 A\cdot\frac{1}{r^3}
  \bigl[4(2\nu-1)-12\nu(\hat r_1^2+\hat r_3^2)
  +60\hat r_1^2\hat r_3^2\bigr].
  \label{eq:gradsum}
\end{equation}

\subsection{Step D: Angular tensor and far-field formula}
\label{sec:stepD}

Step C produced $\partial_3u_1+\partial_1u_3$ as a function of
position; multiplying by $G$ converts it into the physical shear
stress $\sigma_{13}$, and substituting the explicit prefactor
$A=1/[16\pi G(1-\nu)]$ reduces the result to a single radial factor
$1/r^3$ times a dimensionless function of direction only. That
dimensionless function, denoted $\mathcal T_{13}(\hat{\mathbf r})$
below, is the \emph{angular tensor}: it encodes how strongly the
\OSTZ\ promotes shear stress as a function of the direction
$\hat{\mathbf r}$ from the source, independent of distance, and its
sign and zeros (examined in Step E) determine where the \OSTZ\
assists or opposes further slip. Multiplying Equation~\eqref{eq:gradsum}
by $G$ and inserting $A=1/[16\pi G(1-\nu)]$:
\[
  \sigma_{13}=\frac{G\gz V_0}{16\pi(1-\nu)r^3}
  \bigl[-4(2\nu-1)+12\nu(\hat r_1^2+\hat r_3^2)-60\hat r_1^2\hat r_3^2
  \bigr].
\]
Writing this as $G\gz V_0\mathcal T_{13}/[2\pi(1-\nu)r^3]$ requires
dividing the bracket by 8 (ratio of prefactors $2\pi/16\pi=1/8$):
\begin{equation}
  \boxed{\mathcal T_{13}(\hat{\mathbf r})
  =\tfrac{1}{2}-\nu+\tfrac{3\nu}{2}(\hat r_1^2+\hat r_3^2)
  -\tfrac{15}{2}\hat r_1^2\hat r_3^2.}
  \label{eq:Tij}
\end{equation}
\begin{equation}
  \boxed{\sigma_{13}(\mathbf r)=\frac{G\gz V_0}{2\pi(1-\nu)}
  \cdot\frac{\mathcal T_{13}(\hat{\mathbf r})}{r^3}.}
  \label{eq:dipole}
\end{equation}
The $r^{-3}$ decay follows from differentiating the Kelvin $r^{-1}$
Green's function twice: two orders faster than the Volterra $r^{-1}$
field.

\subsection{Step E: Verification in special directions}
\label{sec:stepE}

The checks below confirm the sign pattern of the angular factor in
directions where the answer can be found independently. Where the factor
is positive, a sheared unit assists the same shear in a neighbour; where
it is negative, it opposes it.

The angular factor has both positive and negative sectors. This is
consistent with elastic interactions promoting some rearrangements and
opposing others.

\paragraph{Glide plane ($x_3=0$, $\hat r_3=0$)}
$\mathcal T_{13}|_{\hat r_3=0}=\half-\nu+(3\nu/2)\hat r_1^2$.
Expanding $r^2=x_1^2+x_2^2$ and collecting:
\begin{equation}
  \sigma_{13}(x_1,x_2,0)=\frac{G\gz V_0}{4\pi(1-\nu)r^5}
  [(1+\nu)x_1^2+(1-2\nu)x_2^2].
  \label{eq:glidecheck}
\end{equation}
Positive definite for $0<\nu<\half$.

\paragraph{On-axis ($\hat r_1=0$, $\hat r_3=1$)}
$\mathcal T_{13}=\half-\nu+3\nu/2=(1+\nu)/2>0$. The same leading shear-dipole value holds along $\hat x_1$ by the
$x_1\leftrightarrow x_3$ symmetry of this shear source, not by axial
symmetry of its shape.

\paragraph{Nodal directions}
In the $x_1$--$x_3$ plane, put $\hat r_1=\cos\theta$ and
$\hat r_3=\sin\theta$. Then
$\mathcal T_{13}=(1+\nu)/2-(15/8)\sin^2(2\theta)$.
It vanishes at $\sin^2(2\theta)=4(1+\nu)/15$ and is negative
at $\theta=\pi/4$ for $0<\nu<1/2$. Thus nodal directions exist in
$x_3\ge0$, although the shear contribution is positive on the glide
plane for this range of $\nu$.

The dilatational moment adds, to leading order,
\begin{equation}
 \sigma_{13}^{\rm dil}=-\frac{G(1+\nu)\varepsilon_0V_0}{2\pi(1-\nu)}
 \frac{x_1x_3}{r^5}.
 \label{eq:dilatfar}
\end{equation}
It vanishes on $x_3=0$ but contributes off that plane.

\subsection{Glide-plane stress: six-step derivation}
\label{sec:glideplane}

The same leading shear stress can be obtained directly on the glide
plane by differentiating the full three-dimensional Green tensor before
setting $x_3=0$.

Working at $x_3=0$ ($r^2=x^2+y^2$, $x\equiv x_1$, $y\equiv x_2$).

\paragraph{Step~1 (off-diagonal $\mathcal G_{13}$)}
$\partial^2(\mathcal G_{13})/(\partial x_1\partial x_3)$ at $z=0$:
gives $A(y^2-2x^2)/r^5$ (using $\partial_3(x_1x_3/r^3)|_{z=0}
=x_1/r^3$, then $\partial_1$).

\paragraph{Step~2 (diagonal $\mathcal G_{33}$)}
At $z=0$: $\mathcal G_{33}|_{z=0}=A(3-4\nu)/r$.
$\partial_1^2(1/r)=(3x^2-r^2)/r^5=(2x^2-y^2)/r^5$:
\[
  \partial_1^2\mathcal G_{33}\big|_{z=0}=A(3-4\nu)(2x^2-y^2)/r^5.
\]

\paragraph{Step~3 (diagonal $\mathcal G_{11}$)}
At $z=0$: $\mathcal G_{11}|_{z=0}=A[(3-4\nu)/r+x^2/r^3]$.
$\partial_3^2(1/r)|_{z=0}=-1/r^3$;
$\partial_3^2(1/r^3)|_{z=0}=-3/r^5$:
\[
  \partial_3^2\mathcal G_{11}\big|_{z=0}=A[-(3-4\nu)/r^3-3x^2/r^5].
\]
Expanding: $-(3-4\nu)r^2-3x^2=-(6-4\nu)x^2-(3-4\nu)y^2$.

\paragraph{Step~4 (collect numerator, units of $A/r^5$)}
\begin{align*}
  x^2\text{ coeff.:}&\;2(y^2\!-\!2x^2)\to-4\\
  &+(6\!-\!8\nu)-(3\!-\!4\nu)-3=-4(1+\nu),\\
  y^2\text{ coeff.:}&\;2-(3\!-\!4\nu)-(3\!-\!4\nu)=-4(1-2\nu).
\end{align*}
Numerator: $-4A[(1+\nu)x^2+(1-2\nu)y^2]/r^5$.

\paragraph{Step~5 (assemble)}
\[
  \sigma_{13}(x,y,0)=\frac{G\gz V_0[(1+\nu)x^2+(1-2\nu)y^2]}
  {4\pi(1-\nu)r^5}.
\]

\paragraph{Step~6 (re-express with $\beff$, restore finite $W$)}
Using $\beff=\gz W$:
\begin{align*}
  G\gz V_0&=G\gz\,\tfrac{2\pi W^3}{3}=G\beff\,\tfrac{2\pi W^2}{3},\\
  \text{so}\quad\frac{G\gz V_0}{4\pi(1-\nu)}&=\frac{G\beff W^2}{6(1-\nu)}.
\end{align*}
A phenomenological regularisation $r^2\to r^2+W^2$ gives
\begin{equation}
  \boxed{\sigma_{13}(x,y,0)=
  \frac{G\beff W^2[(1+\nu)x^2+(1-2\nu)y^2]}
  {6(1-\nu)(x^2+y^2+W^2)^{5/2}}.}
  \label{eq:glideplane}
\end{equation}
This surrogate is non-negative for $0<\nu<1/2$ and finite at the
origin, but is not the exact finite-inclusion solution. In particular,
it vanishes at the centre, whereas the interior shear stress is
$-G\beta_1\gz$. It must not be used to infer the actual core stress.

\paragraph{Physical interpretation}
Along $\hat x$ ($y=0$): promotes forward shear.
Along $\hat y$ ($x=0$): promotes lateral shear.
Thus aligned shear events have an attractive leading interaction in
this plane for $0<\nu<1/2$. The off-plane sectors and different
orientations need not have the same interaction sign.

\subsection{Slip-amplitude convention}
\label{sec:beff}
This subsection defines how much slip one unit carries. It is the
quantity that is added up when a row of units forms a dislocation in
Part~B. For a prescribed simple plastic shear, integrating the engineering
shear eigenstrain along the central thickness gives
\begin{equation}
 \beff=\gz W.\label{eq:beff}
\end{equation}
This is a transformation-slip amplitude, not a lattice Burgers vector
or a displacement discontinuity at a coherent inclusion interface.
The actual elastic displacement is continuous. Moreover, symmetric
shear eigenstrain alone does not specify the plastic rotation; the simple
shear convention must be supplied to interpret this integral as slip.

At projected radius $r_\parallel$, the spheroid thickness is
$W\sqrt{1-r_\parallel^2/W^2}$. Averaging the corresponding transformation
slip over the projected disk gives
\begin{equation}
 \langle b_{\rm tr}\rangle_A=\frac{\gz V_0}{\pi W^2}
 =\frac{2}{3}\gz W.
 \label{eq:meanSlip}
\end{equation}
This explains the factor $2/3$ between central slip and moment-equivalent
uniform slip. The ratio remains $2/3$ for every spheroidal aspect ratio;
it does not approach one in the thin-disk limit at fixed eigenstrain.

\section{Properties of a single unit}
\label{sec:single}
This section collects the properties of one unit that follow from the
energy of Section~\ref{sec:ostz} and the stress of Section~\ref{sec:dipole},
before any assumption about many units is made.

\subsection{Elastic work scale}
\label{sec:strength}
This subsection gives the stress scale at which one unit can be
activated by the applied stress alone. The stress at which the work on one unit equals its elastic energy is
\begin{equation}
 \tau_{\rm el}=\frac{E_{\rm el}}{\gz V_0}
 =\frac{G}{2}\left(\beta_1\gz+
             \frac{\beta_2\varepsilon_0^2}{\gz}\right),
 \label{eq:strength}
\end{equation}
which reduces to $\beta_1G\gz/2$ for pure shear. It is an elastic scale;
the ideal shear strength of a crystal, the stress maximum along a relaxed
shear path~\citep{ogata2004}, is a separate quantity. Within transition
state theory with a symmetric barrier (Section~\ref{sec:sinh} of Part~B), the
forward barrier $F^\ddagger-\tau\gz V_0/2$ vanishes at
$\tau=2F^\ddagger/(\gz V_0)$, which bounds the stress range of the
activated description.

\subsection{One inclusion compared with a straight line defect}
\label{sec:nomacroslip}
Table~\ref{tab:compare1} contrasts the elastic properties of one
inclusion with those of a straight dislocation. The leading inclusion
stress, Equation~\eqref{eq:dipole}, falls as $r^{-3}$; normalised at $r=W$,
it is $0.8\%$ at $r=5W$, compared with $20\%$ for an $r^{-1}$ line field
(Figure~S2 in the Supporting Information). A Burgers circuit that does
not pass through a compatible inclusion has zero closure failure. A
circuit that crosses the sheared region, however, encloses its
transformation slip, up to $\beff=\gz W$ at the centre
(Section~\ref{sec:beff}). In this sense a sheared oblate spheroid is
equivalent to a smeared dislocation loop whose Burgers vector is a
fraction of a lattice vector: a partial dislocation loop of non-lattice
Burgers vector, with edge character where the slip direction is normal
to the rim of the unit ($\pm x_1$) and screw character where it is
parallel to it ($\pm x_2$). It becomes a lattice partial only when the
slip of two to three such units adds up to the Burgers vector of a Shockley partial
(Section~\ref{sec:Nc}).

\begin{table}[!htbp]
\centering
\small
\caption{One homogeneous oblate spheroidal inclusion compared with a straight
dislocation in isotropic linear elasticity.}
\label{tab:compare1}
\renewcommand{\arraystretch}{1.3}
\begin{tabularx}{\linewidth}{>{\raggedright\arraybackslash}p{3.3cm}LL}
\toprule
Property & One inclusion & Straight dislocation\\
\midrule
Leading far-field stress & $\propto r^{-3}$ (force dipole) & $\propto r^{-1}$\\
Closure failure (Burgers vector) & 0 outside the unit; up to $\gz W$ for a circuit through it & $\mathbf b$\\
Interior or core stress & Uniform and finite; $-G\beta_1\gz$ for pure shear & Singular in linear elasticity; finite only with a core model\\
Stress at $r=5W$ relative to $r=W$ & $0.008$ & $0.2$\\
Elastic energy & $\tfrac12(\beta_1\gz^2+\beta_2\varepsilon_0^2)GV_0$ (finite) & Per unit length, e.g.\ edge $\frac{Gb^2}{4\pi(1-\nu)}\ln(R/r_0)$\\
Kinetics & Barrier and attempt rate must be supplied & Barrier and attempt rate must be supplied\\
\bottomrule
\end{tabularx}
\renewcommand{\arraystretch}{1}
\end{table}

\parthead{Part B. Connection with dislocation theory, grain boundary sliding and superplastic flow}

Part~B uses the single-unit results of Part~A. It first lists every
input, assumption and proposal that the connection requires. It then
shows how a row of units gives the slip profile of a dislocation, how
quantities of dislocation theory such as the Peierls stress and the
Frank--Read stress follow, and how the unit leads to the flow law of
superplastic deformation at a single boundary and, with further inputs,
to the mesoscopic scale and the inverse Hall--Petch effect.

\section{Inputs, assumptions and proposals}
\label{sec:assumptions}
Table~\ref{tab:assumptions} collects all the inputs, assumptions and
proposals needed to carry the single-unit results of Part~A into
dislocation theory and superplastic flow, so that each result of Part~B
can be traced to what it depends on. Model premises
are taken from the boundary-sliding model; proposals are introduced in
this work; inputs are taken from experiment or from atomistic
calculations; classical results are standard dislocation theory.

The last column states how each entry is justified. Three kinds of
justification are used: (E) a proposition validated by experiment or by
atomistic calculation; (T) a physically realistic number or form based
on theory or on precedent in the literature; and (F) a fitting constant,
chosen so that the theoretical result agrees with experiment. An entry of
type (F) is yet to be predicted by the present analysis.

\begin{table}[!htbp]
\centering
\small
\caption{Proposals and inputs that connect the elastic analysis with
dislocation theory and with the mesoscopic boundary-sliding model.
Basis: (E) validated by experiment or atomistic calculation; (T)
physically realistic, from theory or precedent; (F) fitting constant.}
\label{tab:assumptions}
\renewcommand{\arraystretch}{1.25}
\begin{tabularx}{\linewidth}{>{\raggedright\arraybackslash}p{4.2cm}>{\raggedright\arraybackslash}p{2.8cm}>{\raggedright\arraybackslash}p{2.1cm}L}
\toprule
Proposal or input & Used for & Type & Basis\\
\midrule
Transition state theory for unit events; $F^\ddagger=E_{\rm el}$ & Flow law, Section~\ref{sec:sinh} & Model premise & T \citep{padmanabhan1977,pad-part1}; E, measured rates of a Cu alloy (\ref{app:ensemble})\\
Unit radius $W=2.5b$ & Energies, Section~\ref{sec:ostz} & Input & T, boundary width of 2--3 atomic spacings~\citep{pad-part1}\\
Strains $\gz$, $\varepsilon_0=\gz/\sqrt3$ & Energies, Section~\ref{sec:ostz} & Input & F, fitted for 41 systems~\citep{harisankar2025}; $\varepsilon_0$ from $\gz$ by von Mises (T); free volume of the unit of the order of the excess free volume of high-angle boundaries, about $0.1\Omega$ per atom, from the molecular-dynamics simulations of \citet{wolf1990} (E)\\
Attempt frequency $\nu_0=kT/h$ & Absolute strain rate & Model premise & T, transition state theory\\
Active fraction $\eta$ & Absolute strain rate & Input & Fraction activated at an instant: T, statistical mechanics and transition state theory (Boltzmann factor in Equation~\eqref{eq:rate}); fraction of sites able to act as units, possibly misorientation dependent: F, not reduced with the fitted strains (\ref{app:ensemble})\\
Threshold stress $\tau_0$ & Flow law, mesoscopic scale & Input & F, fitted~\citep{paddivinski2024}; form T~\citep{padgleiter2004}, an athermal (long-range stress) obstacle as in transition state theory~\citep{seeger1958,kocks1975}\\
P1: Lorentzian slip density of width $W$ & Peierls--Nabarro profile & Proposal & T, Peierls--Nabarro form\\
P2: slips of a row add about a common centre & Peierls--Nabarro profile & Proposal & T, linear superposition\\
P3: accumulated slip equals a partial or full $b$ & $N_c$; emission & Proposal & T, lattice periodicity; E, partial emission in MD (\ref{app:emission})\\
Misfit amplitude satisfying Equation~\eqref{eq:taumax} & Consistency of P1 and P2 & Input (crystal) & T, Peierls--Nabarro theory\\
Lattice resistance amplitude and period & Peierls stress & Input (crystal) & T, Peierls--Nabarro theory\\
Segment length $\ell=2W$ & Line energy of a row & Proposal & T, unit diameter\\
Line tension $Gb^2/[4\pi(1-\nu)]$; $L\approx d$ & Frank--Read stress & Classical; proposal & T, classical; source length set by grain size\\
Excess shear energy stored in the fault & Stacking-fault energy & Proposal & T; compared with measured values (E)\\
Measured fault energy & Stacking-fault energy & Input & E; used to fix an effective $\beta_1\approx0.80$ (F)\\
Crossover angle $\theta_m$ & Boundary energy & Input & E, empirical\\
Sign of $J_\Sigma$; $\bar n=\pi W^2\rho$; spacing $\propto\rho^{-1/2}$ & Softening or hardening & Proposal & T, mean field; spacing as in obstacle hardening\\
\bottomrule
\end{tabularx}
\renewcommand{\arraystretch}{1}
\end{table}

In summary, the elastic results of Part~A require no entry of the table.
The proposals P1--P3, the segment length and the interaction proposals
rest on theory or precedent and have not been validated directly; P3 is
supported by the partial-dislocation emission found in the
molecular-dynamics calculations of the Appendix. The flow law uses the
premise of transition state theory, which has been validated against the
measured strain rates of a Cu alloy, together with fitted strains, a
fitted threshold stress and, for absolute rates, a fitted active
fraction. The agreement with experiment in Sections~\ref{sec:validation} and
\ref{sec:meso} is therefore a test of the activation energy of the unit
with fitted strains, not a prediction free of fitting constants.

\section{From a row of units to a planar slip profile}
\label{sec:emergence}
The decay laws of one unit and of a dislocation (Section~\ref{sec:nomacroslip})
are directly related. When many units on
one plane add their slip, the slipped patch, of size $D$, is equivalent
to a dislocation loop around its edge whose Burgers vector equals the
accumulated slip. Close to the edge, for $W\lesssim r\ll D$, the stress
falls as $r^{-1}$, as for a straight dislocation; far from the whole
patch, $r\gg D$, the loop acts as a dipole and the $r^{-3}$ decay
returns. The $r^{-1}$ range therefore grows with the patch, and the
straight dislocation is the limit $D\to\infty$ (Figure~S3 in the Supporting Information). The
question for this section is therefore what a row of units must satisfy
for its accumulated slip to be emitted as a partial or a full
dislocation. Answering it requires three proposals for a row of units on
one plane to carry the slip distribution of a Peierls--Nabarro
dislocation:
\begin{enumerate}
\item[(P1)] the slip of each unit is spread over the plane with a
Lorentzian density of width $W$;
\item[(P2)] the units of a row act about a common centre, so that their
slips add;
\item[(P3)] the accumulated slip equals the Burgers vector of a partial or
full dislocation of the adjoining crystal.
\end{enumerate}
With these proposals the Peierls--Nabarro profile follows, as shown below.

\subsection{Proposal P1: slip density of one unit}
\label{sec:lorentzian}
Proposal P1 states how the slip produced by one unit is spread over the
boundary plane; it is the starting point for adding the slips of many
units. The inclusion stress of Section~\ref{sec:dipole} does not by itself give
a slip profile, because planar elasticity relates slip density to
traction through a nonlocal operator. Proposal P1 introduces the
normalised density
\begin{equation}
 p_W(x,y)=\frac{W}{2\pi(x^2+y^2+W^2)^{3/2}},\qquad
 \iint p_W\,\mathrm dx\,\mathrm dy=1,
 \label{eq:rho2D}
\end{equation}
with units of inverse area. Its projection onto the slip direction is
\begin{equation}
 \rho_1(x)=\beff\int_{-\infty}^{\infty}p_W(x,y)\,\mathrm dy
 =\frac{\beff W}{\pi(x^2+W^2)},\qquad
 \int\rho_1\,\mathrm dx=\beff,
 \label{eq:rho1}
\end{equation}
where $\rho_1$ is the derivative of the disregistry (dimensionless),
not a dislocation line density. The Lorentzian is chosen because it
leads to the Peierls--Nabarro form; its width is identified with the
unit radius $W$. For comparison, with $a^2=x^2+W^2$, the projection of
the regularised glide-plane stress of Equation~\eqref{eq:glideplane} is
\begin{equation}
 \int\sigma_{13}(x,y,0)\,\mathrm dy
 =\frac{G\beff W^2}{6(1-\nu)}
 \left[\frac{4(1+\nu)x^2}{3a^4}+\frac{2(1-2\nu)}{3a^2}\right],
 \label{eq:projectStress}
\end{equation}
which has the same width scale $W$ but is not exactly Lorentzian.
With Fourier convention $\hat\rho_1(k)=\int\rho_1(x)e^{-ikx}\,\mathrm dx$,
\begin{equation}
 \hat\rho_1(k)=\beff e^{-|k|W}.\label{eq:rhohat}
\end{equation}

\subsection{Proposal P2: addition of slip and the arctangent profile}
\label{sec:PN}
\begin{figure}[!htbp]
\centering
\includegraphics[width=.6\linewidth]{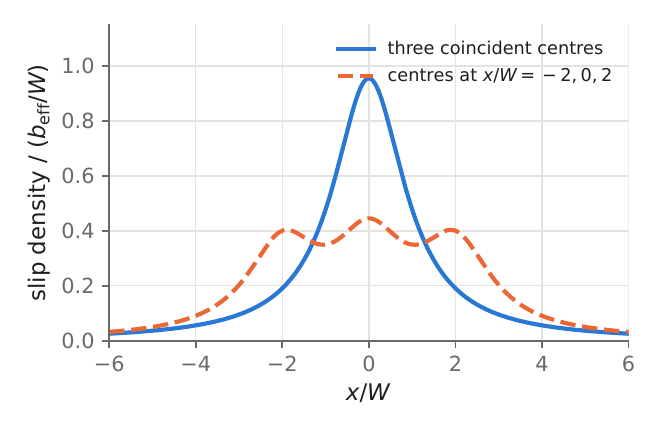}
\caption{Slip density of three Lorentzian profiles with equal total slip:
coincident centres (proposal P2) versus centres at $x/W=-2,0,2$. Only
coincident centres give three times one profile.}
\label{fig:profiles}
\end{figure}

This subsection adds the slips of the units of a row and shows when the
sum has the shape of a Peierls--Nabarro dislocation. For units centred at positions $x_j$, superposition gives
\begin{align}
 \rho_N(x)&=\sum_{j=1}^N\frac{\beff W}{\pi[(x-x_j)^2+W^2]},\\
 U_N(x)&=\sum_{j=1}^N\beff\left[\frac12+
                   \frac1\pi\arctan\frac{x-x_j}{W}\right].
 \label{eq:slip}
\end{align}
Proposal P2 takes the centres to coincide, or their spread to be small
compared with $W$. Separated centres instead give a broader profile with
several peaks (Figure~\ref{fig:profiles}). Under P2,
\begin{equation}
 U(x)=\frac b2+\frac b\pi\arctan\frac{x}{W},\qquad b=N\beff,
 \label{eq:PNprofile}
\end{equation}
which is the Peierls--Nabarro profile with half-width $W$.

In the Peierls--Nabarro model the half-width is fixed by force balance
with the restoring stress of the crystal, $\zeta=Gb/[4\pi(1-\nu)
\tau_{\max}]$ for an edge dislocation~\citep{peierls1940,nabarro1947}.
Identifying $\zeta=W$ therefore requires
\begin{equation}
 \tau_{\max}=\frac{Gb}{4\pi(1-\nu)W},
 \label{eq:taumax}
\end{equation}
which for $W=2.5b$ and $\nu=1/3$ gives $\tau_{\max}\approx G/21$. This
is the condition that a measured or computed misfit surface of the
crystal~\citep{lu2000gsf} must satisfy for P1 and P2 to hold together.

\subsection{Stress of the profile and the Volterra limit}
\label{sec:chainStress}
This subsection checks that the profile obtained under P1 and P2 has the
stress field of a dislocation. For a straight edge dislocation along $x_2$ with planar slip density
$\rho_N$, the glide-plane stress is
\begin{equation}
 \sigma_{13}(x)=\frac{G}{2\pi(1-\nu)}\,\mathrm{P.V.}
 \int_{-\infty}^{\infty}\frac{\rho_N(x')}{x-x'}\,\mathrm dx',
 \label{eq:conv}
\end{equation}
where P.V.\ denotes the Cauchy principal value of the integral. For the
profile of Equation~\eqref{eq:PNprofile},
\begin{equation}
 \sigma_{13}(x)=\frac{Gb}{2\pi(1-\nu)}\frac{x}{x^2+W^2},\qquad
 \max|\sigma_{13}|=\frac{Gb}{4\pi(1-\nu)W}.
 \label{eq:chainStress}
\end{equation}
For $|x|\gg W$ this already has the $1/x$ form of a straight
dislocation, and at fixed $b$
\begin{equation}
 \lim_{W\to0}\sigma_{13}(x)=\frac{Gb}{2\pi(1-\nu)x},
 \label{eq:volterra}
\end{equation}
the Volterra field. This is the edge field of the slipped patch
discussed in Section~\ref{sec:nomacroslip}; at distances larger than a
bounded row itself, the field returns to the $r^{-3}$ form of
Equation~\eqref{eq:dipole}.

\subsection{Proposal P3: number of units per dislocation}
\label{sec:Nc}
This subsection counts how many units are needed to form a partial or a
full dislocation, which decides which of the two a boundary emits. Under
P2 and P3, the number of units whose slip adds up to a Burgers
vector $b$ is
\begin{equation}
 N_c=\frac{b}{\gz W}.\label{eq:Nc}
\end{equation}
For $W=2.5b_{\rm latt}$ and $\gz=0.10$, $N_c=4$ for a full dislocation.
For an FCC Shockley partial, $b_p=b_{\rm latt}/\sqrt3$ and
$N_c=4/\sqrt3\approx2.3$, i.e.\ two to three units. Fewer units are
therefore needed to emit a partial than a full dislocation, which is
consistent with partial-dislocation emission at small grain sizes and
full-dislocation emission at larger ones. $N_c$ is a kinematic count;
whether a row is emitted depends on the energy balance and on the
boundary stress, which enter through the kinetics of
Section~\ref{sec:sinh}.

A set of compatible units carries no lattice Burgers vector. The
lattice dislocation appears at the edge of the slipped patch once the accumulated slip is
accommodated as plastic slip on a plane of the adjoining crystal, which
is the content of P3. This picture agrees with the boundary-sliding model
of \citet{pad-part1}, in which units are associated with free-volume
sites of high-angle boundaries, and with \citet{padgleiter2012}, who
distinguish an elementary sliding event from the lattice-dislocation
accommodation that follows. Boundary emission of dislocations is
observed in atomistic simulations of nanocrystalline
metals~\citep{yamakov2002,yamakov2004,wolf2005}; cooperative sliding of
grain groups~\citep{astanin-part2} is its mesoscopic counterpart.

\section{Estimates of dislocation-related quantities}
\label{sec:derived}
This section uses the energy of Section~\ref{sec:ostz} and the profile of
Section~\ref{sec:emergence} to estimate quantities of dislocation theory.
Each estimate combines the elastic results with a stated proposal and,
where noted, an experimental input; all are listed in
Section~\ref{sec:assumptions}. These derivations show how important
relations of dislocation theory can be obtained from the Eshelby analysis
of a single inclusion; they are not part of the argument for
superplastic flow, which resumes in Section~\ref{sec:gb}.

\subsection{Peierls stress}
\label{sec:peierls}
The Peierls stress is the stress needed to move a dislocation through the
periodic resistance of the lattice. If the dislocation of Section~\ref{sec:emergence} moves through a
periodic lattice resistance, the Lorentzian profile gives a Peierls-type
stress. Let $\mathcal V$ (J\,m$^{-2}$) be the amplitude and $a_P$ the
period of that resistance. Convolving
$\mathcal V[1-\cos(2\pi(x-x_0)/a_P)]$ with the density
$\rho=bW/[\pi(x^2+W^2)]$ gives the energy per unit length
\begin{equation}
 E_{\rm pin}(x_0)\big|_{\rm osc}
 =-\mathcal V b e^{-2\pi W/a_P}\cos(2\pi x_0/a_P),
 \label{eq:Emisfit}
\end{equation}
and the maximum resolved stress
\begin{equation}
 \tau_P=\frac{2\pi\mathcal V}{a_P}e^{-2\pi W/a_P}.
 \label{eq:peierls}
\end{equation}
With $\mathcal V=\tau_{\rm amp}a_P/(2\pi)$ the prefactor is the stress
amplitude $\tau_{\rm amp}$. The reduction factor $e^{-2\pi W/a_P}$ is
$4.3\times10^{-2}$ for $W/a_P=0.5$ and $1.9\times10^{-3}$ for
$W/a_P=1$ (Figure~S6a in the Supporting Information). The amplitude and period are inputs from the
crystal, for example from its generalised stacking-fault
energy~\citep{lu2000gsf}; nonplanar BCC screw cores need a separate core
model~\citep{nabarro1997,xiang2008}.

\subsection{Energy of a row of units and line energy}
\label{sec:coreE}
This subsection compares the energy stored in a row of units with the
core energy of a dislocation. Neglecting interactions, the shear energy of the $N_c$ units of a row is
\begin{equation}
 E_{\rm cluster}=N_c E_{\rm el}^{\rm shear}
 =\frac{\pi\beta_1 b\gz GW^2}{3}.\label{eq:coreE}
\end{equation}
A dislocation core energy is quoted per unit length, $\alpha_cGb^2$.
Assigning the row to a segment of length $\ell=2W$ gives
\begin{equation}
 \frac{E_{\rm cluster}}{\ell}
 =\frac{\pi\beta_1 b\gz GW}{6},\qquad
 \alpha_c=\frac{\pi\beta_1\gz}{6}\frac Wb,
\end{equation}
so that $\alpha_c=0.058$ for $\beta_1=0.4456$, $\gz=0.10$ and $W=2.5b$.
The segment length is an assumption of this estimate.

\subsection{Frank--Read source}
\label{sec:FR}
Once emitted, a dislocation segment pinned at both ends can multiply by
Frank--Read action. For a segment of length $L$ with line tension
$T_\ell$, the critical stress is reached when the bowed segment becomes a
semicircle, $\tau b=2T_\ell/L$~\citep{frank-read1950}. With the classical
estimate $T_\ell\simeq Gb^2/[4\pi(1-\nu)]$,
\begin{equation}
 \tau_{\rm FR}=\frac{2T_\ell}{bL}\simeq\frac{Gb}{2\pi(1-\nu)L}.
 \label{eq:FR}
\end{equation}
The present model supplies the length: if the pinned segment is of the
order of the boundary segment from which it is emitted, $L\approx d$.
With $G=48.3$~GPa, $b=0.2556$~nm and $\nu=1/3$, Equation~\eqref{eq:FR}
gives $2.95$~MPa for $d=1~\mu$m and $0.29$~MPa for $d=10~\mu$m (Figure~S6c
in the Supporting Information). The identification $L\approx d$ is listed
in Section~\ref{sec:assumptions}. It links Frank--Read source action to
the boundary segment from which the dislocation is emitted, a link that
is useful when superplastic flow is derived at the mesoscopic scale
(Section~\ref{sec:meso}).

\subsection{Stacking-fault energy}
\label{sec:SFE}
A Shockley partial formed from a row of units (Section~\ref{sec:Nc})
leaves a stacking fault behind it. This subsection estimates the fault
energy under the proposal that the shear energy of the units in excess
of that carried by the partial is stored in the fault. In an FCC crystal
$|b_{\rm partial}|/|b_{\rm full}|=1/\sqrt3$, and the proposal gives the
energy per unit area
\begin{equation}
 \gamma_{\rm est}=\frac{(\sqrt3-1)E_{\rm el}^{\rm shear}}{\pi W^2}
 =\frac{(\sqrt3-1)\beta_1\gz^2GW}{3}.\label{eq:SFE}
\end{equation}
For Cu with $G=48.3$~GPa, $W=b=0.2556$~nm and $\gz=0.137$,
Equation~\eqref{eq:SFE} gives $25.2$~mJ\,m$^{-2}$ with the Eshelby coefficient
$\beta_1=0.4456$ and $56.5$~mJ\,m$^{-2}$ with an effective boundary
coefficient $\beta_1=1$ (Figure~S6b in the Supporting Information). The measured intrinsic
stacking-fault energy of Cu, about $45$~mJ\,m$^{-2}$~\citep{gallagher1970},
lies between these values; taken as an experimental input, it fixes the
effective coefficient at $\beta_1\approx0.80$. A predictive treatment
would use the generalised stacking-fault energy of the
crystal~\citep{vitek1968,lu2000gsf}. The FCC factor $\sqrt3-1$ applies to
FCC partials; HCP faults and twin boundaries need their own
crystallographic factors.

\section{Grain-boundary energy}
\label{sec:gb}\label{sec:RS}
This section relates the dislocation picture to the energy of the
boundary in which the units sit, which enters the mesoscopic
boundary-sliding analyses~\citep{pad-part1,harisankar2025}. A low-angle
tilt boundary is an array of edge dislocations with spacing
$D\simeq b/\theta$. Including the core in a constant $A$ gives the
Read--Shockley energy~\citep{readshockley1950}
\begin{equation}
 \gamma_{\rm GB}(\theta)=E_0\theta(A-\ln\theta),\qquad
 E_0=\frac{Gb}{4\pi(1-\nu)}.\label{eq:RS}
\end{equation}
Choosing $A=1+\ln\theta_m$ places the maximum at a crossover angle
$\theta_m$, where $\gamma_{\rm GB}=E_0\theta_m$. With the Cu inputs,
$E_0=1.474$~J\,m$^{-2}$, and $\theta_m=15^\circ$ gives
$0.386$~J\,m$^{-2}$ (Figure~S7 in the Supporting Information). The crossover angle is an experimental
input, and the constant energy above it is the usual empirical
continuation. High-angle boundary energies also depend on the boundary
plane; when needed they are taken from experiment or atomistic
calculations, as in the mesoscopic analyses. For glasses the
corresponding quantity is a fitted interface parameter.

\section{Equilibrium populations and elastic interactions}
\label{sec:stat}\label{sec:Hamiltonian}
This section describes how neighbouring units influence one another
through the dipole stress of Section~\ref{sec:dipole}, and whether that
interaction raises or lowers the stress needed to activate them.
Let $n_i\in\{0,1\}$ denote occupation of a transformed state, with
state free-energy difference $E_i$, and set
$h=\tau\gz V_0$. An illustrative lattice-gas Hamiltonian is
\begin{equation}
 \mathcal H=\sum_i(E_i-h)n_i-\frac12\sum_{i\ne j}J_{ij}n_in_j.
 \label{eq:Hamiltonian}
\end{equation}
Here $E_i$ is the free-energy difference between the untransformed and
transformed states; the activation barrier and the threshold stress
enter separately in Section~\ref{sec:sinh}.

For identical aligned pure-shear zones far apart,
\begin{align}
 E_{ij}^{\rm int}&=-\sigma_{13}^{(i)}(\mathbf x_j)\gz V_0=-J_{ij},
 \label{eq:Eint}\\
 J_{ij}&=\frac{G\gz^2V_0^2}{2\pi(1-\nu)r_{ij}^3}
                 \mathcal T_{13}(\hat{\mathbf r}_{ij}).\label{eq:Jij}
\end{align}
Here $\mathcal T_{13}$ is the angular factor of Figure~\ref{fig:schematic}b,
so the interaction is attractive for a neighbour in a positive sector
and repulsive in a negative one. Dilatation and differently oriented
units add further terms, and at separations comparable with $W$ the
finite-inclusion interaction replaces the point-dipole form.

\subsection{Mean field and its sign}
\label{sec:MF}\label{sec:selfconsistency}\label{sec:taylor}
Each unit feels the stress of all its neighbours. The mean-field
approximation replaces these by their average, which shows directly
whether cooperation between units lowers or raises the stress needed to
activate them. Writing $n_i=\bar n+\delta n_i$ and neglecting quadratic fluctuations
gives $n_in_j\simeq\bar n(n_i+n_j)-\bar n^2$. For a homogeneous
representative site, define $J_\Sigma=\sum_{j\ne i}J_{ij}$ and $E_i=E$.
Then
\begin{equation}
 h^{\rm eff}=E-\tau\gz V_0-J_\Sigma\bar n,\qquad
 \bar n=\frac{1}{1+\exp(h^{\rm eff}/kT)}.
 \label{eq:selfconsfull}
\end{equation}
If several solutions exist, equilibrium requires a free-energy stability
comparison; simple iteration need not find the stable solution.
Positive $J_\Sigma$ favours occupation and lowers its stress scale:
\begin{equation}
 \Delta\tau_{\rm MF}=-\frac{J_\Sigma\bar n}{\gz V_0}.
 \label{eq:tau0eff}
\end{equation}
For positive $J_\Sigma$ cooperation lowers the activation stress, that
is, it softens. Hardening requires additional assumptions:
(a) a net negative coupling, $J_\Sigma<0$, which arises when neighbouring
units lie mainly in each other's negative angular sectors
(Figure~\ref{fig:schematic}b); and (b) an identification of the occupation
with a dislocation density, $\bar n=\pi W^2\rho_{\rm disl}$. With the
representative scale $|J_\Sigma|=zG\gz^2V_0^2/[4\pi(1-\nu)W^3]$, these
give
\begin{equation}
 \Delta\tau_{\rm MF}=\mp\frac{z\gz G\pi W^2}{6(1-\nu)}\rho_{\rm disl},
 \label{eq:Taylorhardening}
\end{equation}
with the upper sign for $J_\Sigma>0$ (softening) and the lower sign for
$J_\Sigma<0$ (hardening). (c) The Taylor form $\alpha_TGb\sqrt{\rho_{\rm
disl}}$ further requires the spacing of interacting units to scale as
$\rho_{\rm disl}^{-1/2}$, as in obstacle-controlled hardening. Figure~S5 in the Supporting Information compares the two forms.

\section{Activated flow and disorder}
\label{sec:sinh}\label{sec:unified}
This section gives the flow law of superplastic deformation at a single
boundary. Following \citet{padmanabhan1977} and
\citet{pad-part1}, the rate of unit events is described by transition
state theory, which is adopted here as part of the model. According to the model of Padmanabhan et al., the
oblate-spheroid description is intended for elevated temperatures, at
which this treatment applies.

\subsection{Transition state theory and the sinh law}
Steady flow requires repeated transitions at renewable sites. Let
$F^\ddagger$ be the free energy of activation at zero stress, $v=\gz V_0$ the
work volume of one forward step, and $\nu_0=kT/h$ the transition-state
attempt frequency. Assume equivalent successive
states along a repeating slip coordinate and a symmetric transition
state. For a non-negative dissipative threshold $\tau_0$, define
\begin{equation}
 \tau_d=\operatorname{sgn}(\tau)\max(|\tau|-\tau_0,0),\qquad
 \Delta w=\tau_d v/2.
 \label{eq:driving}
\end{equation}
For positive flow above threshold this reduces to $\tau-\tau_0$.
This threshold is a part of the concept of athermal (long-range stress)
obstacles to shear transmission in transition state theory, introduced
in early work on thermally activated glide~\citep{seeger1958,kocks1975};
at zero threshold
the usual unbiased Eyring law is recovered.
\begin{equation}
 \Gamma^\pm=\nu_0\exp[-(F^\ddagger\mp\Delta w)/(kT)],\qquad
 \frac{\Gamma^+}{\Gamma^-}=e^{\tau_d v/(kT)}.
 \label{eq:Gamma}
\end{equation}
For independent renewably active sites of number density $n_s$,
each event contributes plastic shear volume $v$, giving
\begin{equation}
 \dot\gamma=2n_s v\nu_0e^{-F^\ddagger/(kT)}
                  \sinh\!\left(\frac{\tau_d v}{2kT}\right).
 \label{eq:rateGeneral}
\end{equation}
The site density, orientation average, and accommodation conditions
are physical inputs. An effective grain-boundary activity fraction
$n_sV_0=\eta W/d$ gives
\begin{equation}
 \dot\gamma=\frac{2\eta W\gz\nu_0}{d}e^{-F^\ddagger/(kT)}
             \sinh\!\left(\frac{\tau_d\gz V_0}{2kT}\right).
 \label{eq:rate}
\end{equation}
Here $\eta$ is the fraction of boundary sites that can act as units, an
input of the mesoscopic model that may depend on boundary misorientation;
the fraction of these units that is activated at a given instant
follows from statistical mechanics through the Boltzmann factor of
transition state theory, which is already contained in
Equation~\eqref{eq:rate}. For a homogeneous glass the site density
$n_s$ replaces $\eta/d$. Equation~\eqref{eq:rate} has the
forward--reverse structure of the boundary-sliding model~\citep{pad-part1}.

As in the boundary-sliding model, we identify the free energy
of activation with the elastic energy of the unit, $F^\ddagger=E_{\rm el}$
(Section~\ref{sec:ostz}). This is the model's proposal for the barrier;
the rate then follows from Equation~\eqref{eq:rate} with $\nu_0=kT/h$ and
the active fraction $\eta$ (a function of boundary misorientation) as
inputs.

\subsection{Disorder average}
Sites in a real boundary differ in structure, so their barriers differ.
This subsection shows how a spread of barriers changes the rate. For a
normalised positive-barrier distribution $f(F)$ independent of
$v$ and $\nu_0$,
\begin{equation}
 \dot\gamma=\frac{2\eta W\gz\nu_0}{d}
 \sinh\!\left(\frac{\tau_d\gz V_0}{2kT}\right)
 \int_0^\infty f(F)e^{-F/(kT)}\,\mathrm dF.
 \label{eq:rateDisorder}
\end{equation}
Correlated disorder in size, strain, and barrier requires a joint
average of the entire rate expression. The cumulant expansion is
\begin{equation}
 \ln\langle e^{-F/(kT)}\rangle
 =-\frac{\bar F}{kT}+\frac{\sigma_F^2}{2(kT)^2}
 -\frac{\kappa_3}{6(kT)^3}+\cdots.
 \label{eq:rateUnified}
\end{equation}
Truncation after the variance is exact for an untruncated Gaussian,
which allows negative barriers, and only an approximation for a
positive or lognormal distribution. It is not controlled at arbitrarily
low temperature. A lognormal distribution has support $(0,\infty)$;
any quoted finite interval must be defined as a quantile interval or
an explicit truncation.

\subsection{Cooperativity and crystalline flow}
Units may also act in groups. This subsection states what must be
specified before such cooperative events can enter the rate law. A
cluster weighting requires a specified set of states, populations,
and transition rates. If cluster free energies $E_c(N)$ are supplied,
a normalised conditional size distribution and mean are
\begin{equation}
 p_N=\frac{e^{-E_c(N)/(kT)}}{\sum_{m=1}^{N_{\max}}e^{-E_c(m)/(kT)}},
 \qquad \langle N\rangle=\sum_{N=1}^{N_{\max}}Np_N.
 \label{eq:Theta}
\end{equation}
Equal energies give $(N_{\max}+1)/2$, not $N_{\max}$.
Only concentration at the largest size gives that latter mean.
Even $\langle N\rangle$ is not automatically a multiplicative rate
factor: cluster rates, site conservation and active densities must
also be specified. The Orowan relation $\dot\gamma=\rho_m b v_d$
requires a mobile line density $\rho_m$ and dislocation velocity $v_d$.
No such relation follows from an unnormalised cluster sum.

\section{Comparison with experiment and with fitted superplastic-flow parameters}
\label{sec:validation}\label{sec:harisankar}
This section confronts the model with superplastic data in two ways: an
absolute strain rate from the flow law, and the parameters fitted by the
mesoscopic analysis of \citet{harisankar2025} for 41 systems. Further
numerical checks are given in the Supporting Information.

\paragraph{Strain rate} For $T=773$~K, $d=10~\mu$m, $\tau_d=5$~MPa and
the inputs of Section~\ref{sec:ostz}, the bias is $0.001379$~eV and
$\Delta w/kT=0.02070$. With $F^\ddagger=0.40512$~eV,
\begin{equation}
 \dot\gamma=(7.0916\times10^{-10})\eta\nu_0.
 \label{eq:rateExample}
\end{equation}
With the transition-state frequency $\nu_0=kT/h=1.61\times10^{13}$~s$^{-1}$
this is $\dot\gamma=1.14\times10^{4}\,\eta$~s$^{-1}$. The superplastic
range $10^{-3}$--$10^{-1}$~s$^{-1}$ is reached for
$\eta\approx9\times10^{-8}$--$9\times10^{-6}$, i.e.\ when roughly one
boundary site in $10^5$--$10^7$ is active at a time (Figure~S4 in the Supporting Information). This small
fraction reflects the low illustrative barrier of $0.405$~eV. With the
strains fitted to superplastic data, read in the tensor convention of the
boundary-sliding model, and $G=24$~GPa (the boundary shear modulus
computed for a high-angle boundary in Cu by \citet{neier2020}; the shear
modulus can decrease to such low values in nanostructured materials, as
reported experimentally by Weller, Gleiter, Birringer and
others~\citep{gleiter1989,weller1991}), the unit energy is
$1.6$--$1.8$~eV, and the rate law then reproduces measured superplastic
strain rates of a Cu alloy without a reduced active fraction
(\ref{app:ensemble}). A shear rate is converted to a tensile rate
with the usual von Mises factor.

\paragraph{Fitted parameters} For the 41 systems listed in the Supporting Information, the
mean fitted strains are $\bar\gamma_0=0.0851$ and
$\bar\varepsilon_0=0.0493$, with standard deviations $0.0205$ and
$0.0122$. The fitted $\gamma_0$ is a tensor shear strain, linked to the
fitted dilatation by $\varepsilon_0=\gamma_0/\sqrt3$~\citep{harisankar2025};
its mean corresponds to an engineering shear strain of $0.170$, above the
illustrative value $0.10$ used in Section~\ref{sec:ostz}, while the model
dilatation $0.05$ lies $1.5\%$ above the fitted mean. The fitted activation energies $Q$
correlate with the elastic energy of one unit (Pearson $r=0.77$), with a
mean ratio $Q/E_{\rm el}=4.05\pm1.41$ when the fitted strains are treated
as engineering strains. The fitted strains are tensor shear strains;
with that convention the elastic estimate rises by a factor of $2.80$ for
$\varepsilon_0=\gamma_0/\sqrt3$, and the ratio becomes $1.45\pm0.50$. The ratio of fitted to estimated boundary
energies has median $4.45$ and range $0.91$--$48$. Fits of earlier flow
models~\citep{sripathi2014,buenz2015,pad2018} provide further context.

\section{From a single boundary to mesoscopic superplastic flow}
\label{sec:meso}
The analysis so far is confined to a single grain boundary. This section
sets out how it can be carried to superplastic deformation of a
polycrystal, with all its features up to the inverse Hall--Petch effect.
The route has three steps. First, emission from one boundary is followed
to partial dislocations at the lower end of the nanocrystalline
grain-size range and to full dislocations at larger grain sizes, using
the counts of Section~\ref{sec:Nc}. Second, the description is carried to
the mesoscopic scale of grain ensembles, using Herring's analysis of
equilibrium at boundary junctions as done by~\citet{padgleiter2004}.
Third, at the lower end of the nanocrystalline range it is used to
describe the inverse Hall--Petch relation of
\citet{hahnpadmanabhan1997}. Alternatively, the unit-level results can
be merged with the existing mesoscopic analysis
\citep{pad-part1,sripathi2014,paddivinski2024,harisankar2025} by a
one-to-one correspondence of their parameters ($W$, $\gz$,
$\varepsilon_0$, $F^\ddagger$, $\eta$). \ref{app:extension} carries out
each step for Cu with molecular-dynamics calculations, and
\ref{app:synthesis} collects the correspondence.

These calculations have the well-known limitations of molecular-dynamics
simulation. The strain rates that can be simulated, $10^8$--$10^9$~s$^{-1}$,
exceed those of superplastic flow, about $10^{-5}$--$10^{-1}$~s$^{-1}$, by
nine to fourteen orders of magnitude. Embedded-atom potentials are fitted
mainly to properties at 0~K, and their temperature dependence is not
known. Reliable
potentials exist mainly for pure metals, whereas superplastic materials
are microduplex and other complex alloys, ceramics, composites and
metallic glasses; interpreting their behaviour by simulation would need
large advances in methods and potentials. A simulation also contains at
most a few million atoms (up to $2.9\times10^6$ here), against about
$10^{23}$ in an engineering specimen.

The way of connecting the earlier phenomenological models of
superplastic flow \citep{pad-part1,hahnpadmanabhan1997,padgleiter2004,
harisankar2025} with the analysis of a single unit has now been
demonstrated. In view of the present limitations of molecular-dynamics
simulation, future analyses of superplastic flow can therefore use the
phenomenological approach, with the unit-level parameters of Part~A as
its physical basis.

\section{Discussion}
\label{sec:discussion}\label{sec:fundamental}\label{sec:classes}
The analysis gives three kinds of results. Derived from elasticity in
Part~A: the energy coefficients of the unit, its force-dipole stress, and
the sign of the interaction between units. Obtained in Part~B with the
proposals P1--P3: Peierls--Nabarro profile of a row of units and the
number of units per partial or full dislocation. Estimated in Part~B with
experimental inputs: the Peierls stress, the Frank--Read stress, the
stacking-fault energy, the boundary energy and the absolute strain rate.
Atomistic calculations could test the inputs directly: the radius and
eigenstrains of a unit, its transition pathway and barrier, and the
disregistry of a row during emission.

For glasses, polymers and nanocrystals, the interactions, structural
relaxation and accommodation of units differ, and each class requires
its own inputs. Extensions to anisotropic elasticity, finite boundaries
and several unit orientations are needed for quantitative,
material-specific predictions for those cases.

\section{Conclusions}
\label{sec:conclusions}
\paragraph{Part A: a single unit}
\begin{enumerate}
\item \textbf{Elastic energy.} For an oblate-spheroidal unit
($\alpha=1/2$, $\nu=1/3$) the shear and volumetric energy coefficients
are 0.445599 and 0.888889. Within transition state theory this energy
sets the free energy of activation of the unit.

\item \textbf{Stress field.} The shear stress of one unit is a force
dipole that decays as $r^{-3}$ and changes sign with direction, so
neighbours can assist or oppose one another. A sheared unit is equivalent
to a smeared partial dislocation loop of non-lattice Burgers vector.
\end{enumerate}

\paragraph{Part B: dislocation theory and superplastic flow}
\begin{enumerate}
\setcounter{enumi}{2}
\item \textbf{Connection to a dislocation.} With proposals P1--P3, a row
of units has the Peierls--Nabarro profile of half-width $W$. About four
units make a full dislocation and two to three a Shockley partial. A
slipped patch formed by many units has an $r^{-1}$ field near its edge,
like a dislocation.

\item \textbf{Dislocation-related estimates.} With stated inputs the
model gives a Peierls stress, a Frank--Read stress of $0.3$--$3$~MPa for
$d=1$--$10~\mu$m, and stacking-fault energies of $25$--$57$~mJ\,m$^{-2}$
that bracket the measured value for Cu.

\item \textbf{Kinetics.} Transition state theory gives a sinh flow law.
With the fitted strains of a Cu-based superplastic alloy in the tensor
convention, the unit energy agrees with the fitted activation energies to
within about $3\%$ and reproduces all 52 measured strain rates within a
factor of ten (median ratio $0.87$; order of magnitude accuracy).

\item \textbf{Mesoscopic scale.} The route from a single boundary to
mesoscopic superplastic flow and the inverse Hall--Petch regime is set
out in Section~\ref{sec:meso}. Molecular-dynamics calculations for Cu
(\ref{app:extension}) give an activation volume of $2.7$--$5.2b^3$ for
nanocrystalline flow, close to the work volume of one unit, and a
strength maximum consistent with the threshold law of the
boundary-sliding model. Given the limitations of such simulations, the
phenomenological approach, based on the unit-level parameters, remains
the practical route for future analyses until significant advances are
achieved in molecular-dynamics simulation techniques and in the
determination of interatomic potentials, particularly for alloys.
\end{enumerate}

\appendix
\setcounter{figure}{0}
\setcounter{table}{0}
\section{Extension to the mesoscopic scale: molecular-dynamics calculations for Cu}
\label{app:extension}
This Appendix examines how the single-boundary analysis carries over to
dislocation emission, to grain ensembles and to the nanocrystalline range,
and how it connects with the existing mesoscopic boundary-sliding analysis.
Each part states the behaviour expected from the unit analysis, compares it
with a calculation for Cu, and notes what remains to be checked by
experiment or simulation. The calculations use the embedded-atom
potential of \citet{mishin2001} in LAMMPS~\citep{thompson2022}; structures
were built with Atomsk~\citep{hirel2015} and analysed with
OVITO~\citep{stukowski2010}, including dislocation extraction
(DXA)~\citep{stukowski2012}. They are single-material calculations,
intended to make the expected behaviour concrete rather than to settle it;
their limitations are discussed in Section~\ref{sec:meso}. Simulation settings, the stress--strain curves and a list of all runs are
given in Section~S7 of the Supporting Information; inputs, scripts and
processed data are available at \url{https://github.com/albert-hzbn/oblate-spheroid-gbs}.

\subsection{Dislocation emission from a boundary}
\label{app:emission}
\begin{figure}[!htbp]
\centering
\includegraphics[width=.6\linewidth]{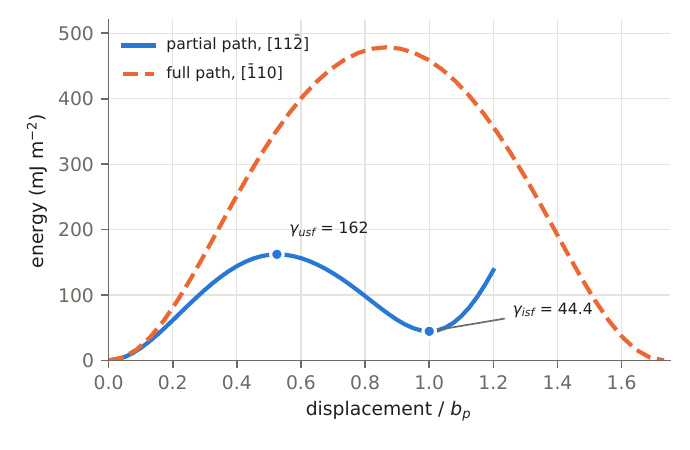}
\caption{Generalised stacking-fault energy of Cu (EAM) on $\{111\}$ along
the partial ($[11\bar2]$) and full ($[\bar110]$) directions, with the
displacement in units of the partial Burgers vector $b_p$.}
\label{fig:A1}
\end{figure}

Section~\ref{sec:Nc} suggests that a row of $N_c$ units emits a Shockley
partial when $N_c\approx2$--$3$ and a full dislocation when $N_c\approx4$,
the choice between the two being decided by the generalised stacking-fault
energy of the receiving crystal and by grain size. The generalised
stacking-fault energy of Cu on $\{111\}$ (Figure~\ref{fig:A1}) gives an intrinsic fault
energy $\gamma_{\rm isf}=44.4$~mJ\,m$^{-2}$, close to the measured
$45$~mJ\,m$^{-2}$~\citep{gallagher1970}, an unstable fault energy
$\gamma_{\rm usf}=162$~mJ\,m$^{-2}$ along the partial path, and
$479$~mJ\,m$^{-2}$ along the full-dislocation path. The low ratio
$\gamma_{\rm isf}/\gamma_{\rm usf}=0.27$ favours partial emission
\citep{vanswygenhoven2004}. With the restoring stress $\pi\gamma_{\rm
us}/b$, force balance gives a Peierls--Nabarro half-width of $0.25$~nm
for both partial and full dislocations, about one Burgers vector; within
the crystal the unit radius is therefore closer to $W\approx b$ than to
the boundary value $2.5a_0$.

\begin{figure}[!htbp]
\centering
\includegraphics[width=\linewidth]{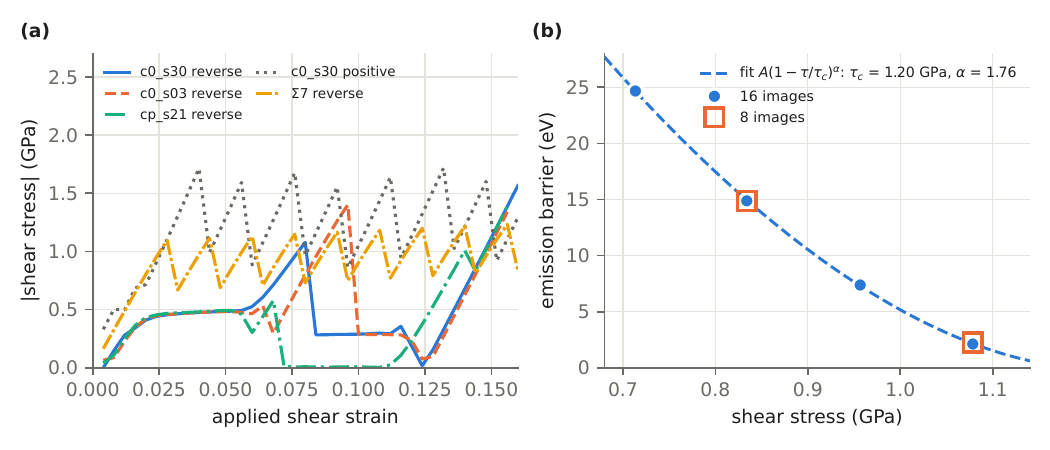}
\caption{(a) Shear stress under 0~K quasi-static shear for three
$\Sigma11$ Cu boundary structures (reverse shear), one of them under
positive shear, and the $\Sigma7$ boundary. (b) Emission barrier versus
applied stress from climbing-image nudged elastic band calculations with
sixteen images (filled) and eight images (open), with a power-law fit.}
\label{fig:A4}
\end{figure}

Emission was examined by shearing periodic bicrystals quasi-statically at
0~K in steps of $0.004$ strain, for three $\Sigma11(557)/(77\bar1)$
boundary structures and the $\Sigma7(415)$ boundary, each in both shear
senses (Figure~\ref{fig:A4}a). Emission occurs only under reverse shear of the
$\Sigma11$ boundaries, at $1.08$, $0.53$ and $0.46$~GPa for the three
structures; a fourth, independently prepared structure relaxed to the first
and reproduced its result exactly. In each case the boundary releases
Shockley partials it already contains: the faulted area grows by
$2\times10^3$--$8\times10^3$ atoms while the dislocation line length does
not increase. Under positive shear the same boundaries reach
$1.7$~GPa and then relax by boundary sliding without fault growth, and the
$\Sigma7$ boundary slides in a stick--slip manner at $0.7$--$1.2$~GPa without
emitting at all. Emission therefore depends on boundary structure and on the
sense of shear.

For the boundary with the highest emission stress, climbing-image nudged
elastic band calculations~\citep{henkelman2000} with sixteen images give
barriers of $2.12$, $7.38$, $14.88$ and $24.67$~eV at $1.08$, $0.96$, $0.83$
and $0.71$~GPa (Figure~\ref{fig:A4}b); eight-image calculations at two of
these stresses agree to within $0.14$~eV. The barrier follows
$A(1-\tau/\tau_c)^{\alpha}$ with $\tau_c=1.20$~GPa and $\alpha=1.76$, and the
local activation volume is $410$--$780b^3$. At 0~K, emission from this
boundary is thus a collective event far larger than one unit.

\begin{figure}[!htbp]
\centering
\includegraphics[width=\linewidth]{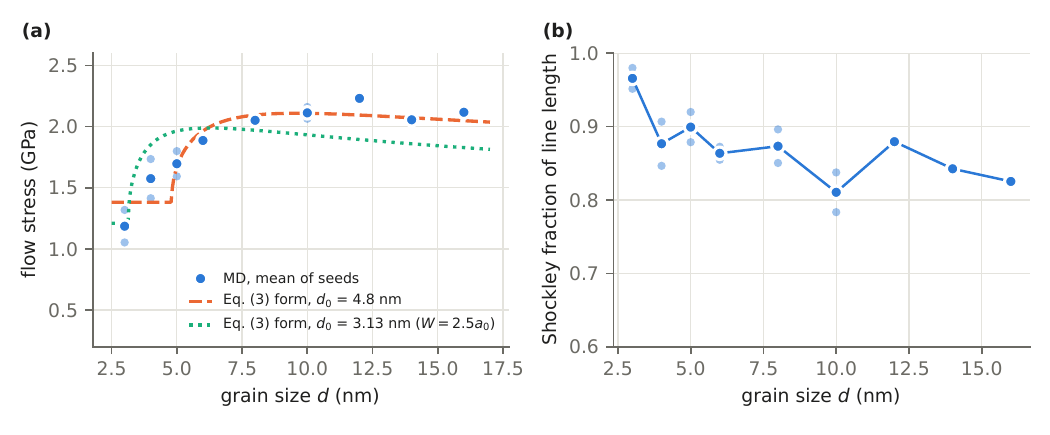}
\caption{Nanocrystalline Cu, 300~K, $5\times10^8$~s$^{-1}$. (a) Flow stress
(mean over 6--10\% strain) versus grain size, with the threshold form of
\citet{hahnpadmanabhan1997} fitted with free $d_0$ and with
$d_0=2\sqrt6\times2.5a_0$. (b) Fraction of dislocation line length
carried by Shockley partials at 10\% strain. Light markers are individual
seeds.}
\label{fig:A2}
\end{figure}

Grain-size dependence was examined by tensile tests at 300~K and
$5\times10^8$~s$^{-1}$ on 16-grain random polycrystals with grain sizes
$d=3$--$16$~nm (up to $2.9\times10^6$ atoms; two seeds for
$d\le10$~nm). At 10\% strain, Shockley partials carry $97\%$ of the
dislocation line length at $d=3$~nm, falling to about $83\%$ at
$14$--$16$~nm (Figure~\ref{fig:A2}b): full dislocations become more frequent
with larger grain size, but partials still dominate up to $16$~nm.

Two questions remain open: the grain size at which full dislocations
become the majority, and whether thermally activated emission at finite
temperature proceeds through events smaller than the 0~K collective
release.

The 0~K calculations do not describe grain boundary sliding itself,
which in the boundary-sliding model is a thermally activated process at
elevated temperatures. They show which boundary structures can emit
dislocations, the stress required without thermal assistance, and the
size of the event that thermal activation must overcome. The
finite-temperature calculations of \ref{app:synthesis}, at 300--700~K
(up to about half the melting temperature of Cu), are those that bear on
the model of superplastic flow.

\subsection{Strain rates of a grain ensemble}
\label{app:ensemble}
\begin{figure}[!htbp]
\centering
\includegraphics[width=.6\linewidth]{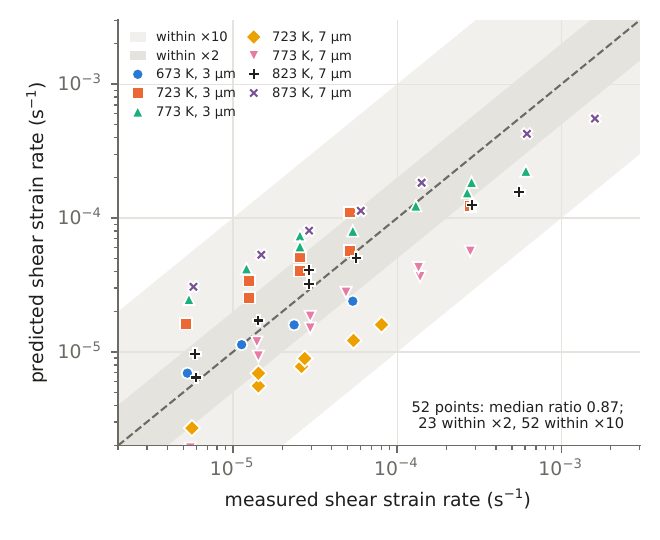}
\caption{Shear strain rates of a Cu-based superplastic alloy predicted
with the rate law of \citet{paddivinski2024} and the unit activation energy
of this work, against the measured rates at all 52 experimental points
(digitised stresses). Shaded bands mark agreement within factors of 2 and
10.}
\label{fig:A5}
\end{figure}

For a grain ensemble, the natural route is to average the rate law of the
boundary-sliding model, with a threshold stress set by accommodation at
triple junctions through Herring's analysis~\citep{padgleiter2004}, over
the grain-size distribution of the material, taking the activation energy
from the unit (Section~\ref{sec:ostz}). As a first check, we test this
activation energy against measured strain rates. The strain $\gamma_0$ of the
mesoscopic model is a tensor shear strain, since its relation to the
uniaxial strain, $\varepsilon_0=\gamma_0/\sqrt3$, follows from the tensorial
von Mises treatment. With this convention the present Eshelby
coefficients give, for the fitted strains of the Cu-based superplastic
alloy of \citet{paddivinski2024}, activation energies of
$158$--$178$~kJ\,mol$^{-1}$, within about $3\%$ of the fitted
$163$--$180$~kJ\,mol$^{-1}$. We inserted these energies, in place of the
fitted ones, into the rate law of that paper, keeping its fitted thresholds
and strains, and evaluated it at the shear stresses of all 52 experimental
points of its Figure~A1 (digitised; the lowest stress of each set agrees with
the tabulated minimum to within $1\%$). The predicted rates have a median
ratio of $0.87$ to the measured ones; all 52 lie within a factor of 10 and
23 within a factor of 2 (Figure~\ref{fig:A5}). The unit-level energy thus
reproduces the measured superplastic strain rates of this alloy without
adjusting the activation energy.

The same comparison for other superplastic alloys, with the threshold
stress computed from boundary-junction accommodation rather than taken from
fits, would test the energy more widely; this is left to a follow-up
paper.

\subsection{The nanocrystalline range and inverse Hall--Petch behaviour}
\label{app:ihp}
The threshold stress
$\tau_0=G[(d_1/d)(1-d_0/d)]^{1/2}$ with $d_0=2\sqrt6W$ of
\citet{hahnpadmanabhan1997} and \citet{paddivinski2024} predicts a
strength maximum at $d=2d_0$ set by the boundary width $W$. The flow
stress of the polycrystals of \ref{app:emission} rises from $1.19$~GPa at $d=3$~nm to a maximum of $2.23$~GPa at about
$12$~nm and then decreases (Figure~\ref{fig:A2}a), in line with earlier
simulations of Cu~\citep{schiotz1998,schiotz2003}. Fitting the form
$A[(1-d_0/d)/d]^{1/2}+B$ gives $d_0=4.8\pm0.2$~nm (maximum at $9.6$~nm)
with a root-mean-square deviation of $0.14$~GPa, against $0.22$~GPa for
$d_0=3.13$~nm ($W=2.5a_0$). The functional form is therefore supported,
with an effective boundary width $W\approx1.0$~nm.

The analysis of \citet{hahnpadmanabhan1997} and the Shockley fraction of
Figure~\ref{fig:A2}b describe two sides of the same grain-size
dependence. In that analysis, flow at small grain sizes is controlled by
boundary sliding, whose threshold depends on grain size through
$d_0=2\sqrt6W$, and the strength maximum marks the grain size below which
sliding of the boundaries, rather than glide of dislocations inside the
grains, dominates. In the present picture the boundary units also emit
dislocations, and the count $N_c$ of Section~\ref{sec:Nc} decides whether
a partial (two to three units) or a full dislocation (about four units)
is emitted; the Shockley fraction measures which of the two prevails. It
is $0.95$--$0.98$ at $d=3$~nm, where nearly all dislocation content is
partial, and $0.8$--$0.9$ between $4$ and $16$~nm, with no marked change
at the strength maximum. In these simulations the maximum is therefore
not caused by a change from partial to full emission. It is consistent
with the threshold of \citet{hahnpadmanabhan1997}, set by the boundary
width, with partial emission from the boundaries as the dominant
dislocation process on both sides of it.

The position of the strength maximum in materials with a known boundary
width would test the predicted dependence on $W$.

\subsection{Activation parameters and correspondence with the mesoscopic model}
\label{app:synthesis}
\begin{figure}[!htbp]
\centering
\includegraphics[width=\linewidth]{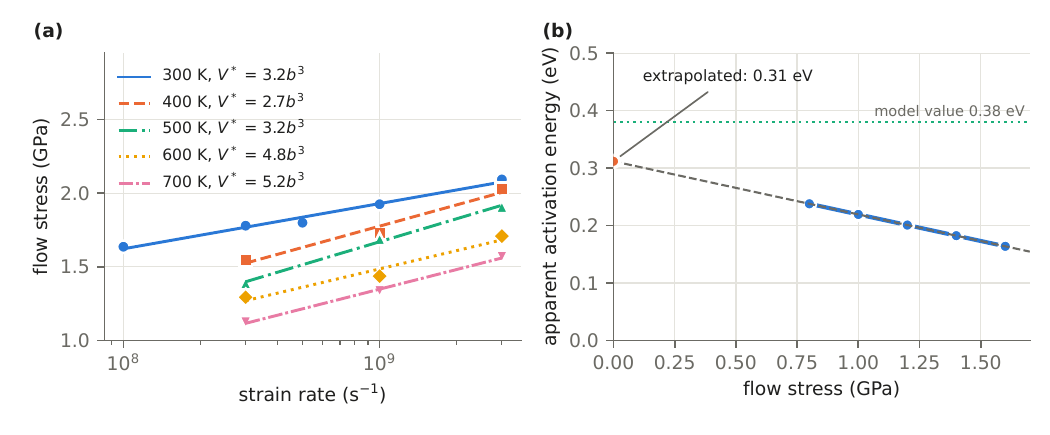}
\caption{Nanocrystalline Cu, $d=5$~nm. (a) Flow stress versus strain rate
at five temperatures with the resulting activation volumes. (b) Apparent
activation energy at fixed flow stress, with its linear extrapolation to
zero stress.}
\label{fig:A3}
\end{figure}

The unit-level results can be connected with the mesoscopic analysis
\citep{pad-part1,sripathi2014,paddivinski2024,harisankar2025} through a
one-to-one correspondence of parameters, for which the activation
parameters of steady state flow are needed. Strain-rate jumps ($10^8$--$3\times10^9$~s$^{-1}$)
at 300, 400, 500, 600 and 700~K for $d=5$~nm (Figure~\ref{fig:A3}) give an
apparent activation volume
$V^*=\sqrt3kT/(\partial\sigma/\partial\ln\dot\varepsilon)$ of
$2.7$--$5.2b^3$, rising slowly with temperature and comparable with the work
volume of one unit, $\gamma_0V_0=3.3b^3$. It is two orders of magnitude
smaller than that of the 0~K collective emission (\ref{app:emission}), so steady state flow of
the nanocrystal is controlled by a much smaller, unit-sized event. The
apparent activation energy at fixed stress decreases from $0.24$ to
$0.16$~eV between $0.8$ and $1.6$~GPa and extrapolates to $0.31$~eV at zero
stress, comparable with the $0.38$~eV of the boundary-sliding model. Table~\ref{tab:A1} collects the
correspondence.

\begin{table}[!htbp]
\centering
\small
\caption{Correspondence between the present unit analysis, the first
calculations of this Appendix and the mesoscopic boundary-sliding model.}
\label{tab:A1}
\renewcommand{\arraystretch}{1.25}
\begin{tabularx}{\linewidth}{>{\raggedright\arraybackslash}p{3.3cm}LL}
\toprule
Quantity & This work & Mesoscopic model\\
\midrule
Shear coefficient & $4\beta_1=1.782$ for the tensor shear strain & $1.779$; agreement within $0.2\%$\\
Activation energy & $158$--$178$~kJ\,mol$^{-1}$ (Eshelby, fitted strains); $0.31$~eV from MD & $163$--$180$~kJ\,mol$^{-1}$ (fits); $0.38$~eV\\
Activation volume & $2.7$--$5.2b^3$ (MD flow); $\gamma_0V_0=3.3b^3$; $410$--$780b^3$ (0~K emission) & ---\\
Boundary width & $W\approx1.0$~nm (fit to MD maximum) & $W=2.5a_0=0.64$~nm\\
Partial or full emission & $N_c=2$--$3$ or $4$; MD partial fraction $0.97\to0.83$ ($d=3\to16$~nm) & partials at small, full at large grain size\\
Strain rate & median ratio $0.87$; 52 of 52 points within $\times10$ & measured\\
\bottomrule
\end{tabularx}
\renewcommand{\arraystretch}{1}
\end{table}

In summary, the calculations for Cu are consistent with the unit
analysis at each scale considered. The Eshelby energy of the unit, with the
fitted strains read as tensor strains, reproduces the measured strain rates
of a superplastic Cu alloy without adjusting the activation energy; the
simulated flow stress shows the strength maximum expected from the
threshold form, with an effective boundary width of about 1~nm;
and steady state flow of the nanocrystal is controlled by an event of the size of
one unit, whereas emission from a boundary at 0~K is a much larger
collective event. These results rest on one material, one interatomic
potential, a limited set of boundaries and, for the strain rates, one alloy
with fitted thresholds and strains, so they support the extension rather
than establish it.

\section*{CRediT authorship contribution statement}
\textbf{A.~Linda:} Methodology, Formal analysis, Investigation (analysis
and computations), Software, Visualization, Writing -- original draft.
\textbf{K.A.~Padmanabhan:} Conceptualization (physical description of
the problem), Supervision, Writing -- review \& editing.

\section*{Funding}
This research did not receive any specific grant from funding agencies in
the public, commercial, or not-for-profit sectors.

\section*{Declaration of competing interest}
The authors declare that they have no known competing financial
interests or personal relationships that could have appeared to
influence the work reported in this paper.

\section*{Declaration of generative AI and AI-assisted technologies in the manuscript preparation process}
During the preparation of this work the authors used Claude (Anthropic)
to assist with numerical checks, the set-up and analysis of the
molecular-dynamics calculations of \ref{app:extension}, figure
preparation and language editing.
After using this tool, the authors reviewed and edited the content as
needed and take full responsibility for the content of the published
article.

\section*{Data availability}
The code, simulation inputs, starting structures and processed data needed
to reproduce all results and figures are available at
\url{https://github.com/albert-hzbn/oblate-spheroid-gbs}. Raw molecular-dynamics trajectories are available from the authors
on request.

\bibliographystyle{elsarticle-harv}
\bibliography{refs}

\clearpage
\setcounter{section}{0}\setcounter{subsection}{0}
\setcounter{figure}{0}\setcounter{table}{0}
\renewcommand{\thesection}{S\arabic{section}}
\renewcommand{\thesubsection}{\thesection.\arabic{subsection}}
\renewcommand{\thefigure}{S\arabic{figure}}
\renewcommand{\thetable}{S\arabic{table}}
\begin{center}
{\Large\bfseries Supporting Information}\par\medskip
{\large Implications of an analysis of the shear transformation of a zone of an oblate spheroid for dislocation theory, grain boundary sliding and superplasticity}\par\medskip
Albert Linda and K.A. Padmanabhan
\end{center}
\bigskip

This supplement follows the two parts of the main text. Sections~S1 and
S2 support Part~A (the single unit; Section~S2 also treats a row of
units), Sections~S3--S6 support Part~B (the connection with dislocation
theory and superplastic flow), and Section~S7 gives the simulation
details of Appendix~A. It has three
purposes: to give the numerical values behind the derived results of the
main text; to show the figures behind its estimates, with the
experimental inputs they use; and to set out, item by item, the
correspondence between the present single-boundary analysis and
the mesoscopic boundary-sliding analyses of \citet{pad-part1},
\citet{sripathi2014} and \citet{harisankar2025}. As in the
oblate-spheroid model of \citet{pad-part1}, transition state theory is taken as part of the
model throughout. Eshelby's analysis of an ellipsoidal inclusion, of
which the oblate spheroid is one among a few shapes considered, is purely elastic; the
connection of the shear of the oblate spheroid with transition state
theory was introduced by \citet{padmanabhan1977} and
\citet{pad-part1}.

\section{Elastic energy of the unit}
The elastic energy of one unit sets the scale of its activation energy.
This section gives the numbers behind the energy coefficients of the main
text, so that each value can be checked independently. For $a_1=a_2=W$, $a_3=W/2$ and $\nu=1/3$, direct quadrature and the
closed form give $I_1=2.97068824$, $I_3=6.62499414$ and
$W^2I_{13}=4.87240787$. Thus $S_{1313}=0.27720028$,
$\beta_1=0.44559943$ and $\beta_2=0.88888889$. The sphere limit is
$S_{1313}=7/30$ and the thin-disk limit is $1/2$. For a homogeneous
isotropic inclusion $S_{iikk}=(1+\nu)/(1-\nu)$, independent of shape
(checked at three aspect ratios), so the volumetric energy of the oblate
unit equals that of a sphere of the same volume. The dependence of
$S_{1313}$ and $\beta_1$ on aspect ratio is shown in Figure~2 of the main
text.

For $W=0.75$~nm, $G=22$~GPa, $\g=0.10$ and $\eps=0.05$,
\begin{align*}
 V_0&=0.88357293~\mathrm{nm}^3,\\
 E_{\rm el}&=0.40512141~\mathrm{eV},\\
 E_{\rm dil}/E_{\rm el}&=0.33275684.
\end{align*}
Within transition state theory, as in the oblate-spheroid model, this
energy is taken as the free energy of activation of the unit,
$F^\ddagger=E_{\rm el}$. It is $6.61\%$ above the value $0.38$~eV of
the boundary-sliding model. The added free volume is
$0.04417865$~nm$^3$ per unit, or $0.05\Omega$ per atom if the unit
contains $V_0/\Omega$ atoms each of volume $\Omega$.

\section{Stress field of one unit and of a row of units}
\begin{figure}[!htbp]
\centering
\includegraphics[width=.6\linewidth]{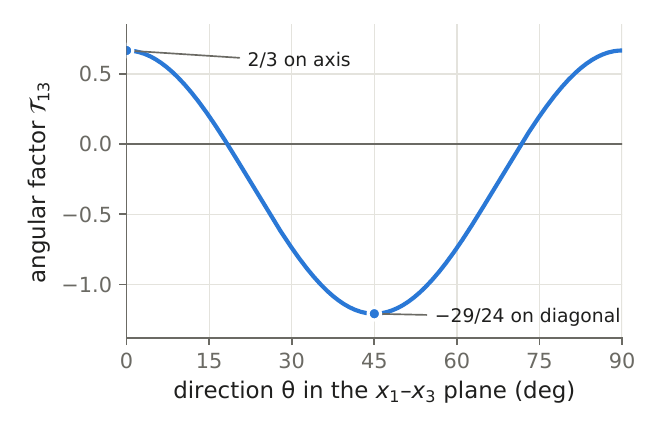}
\caption{Angular factor of the leading shear stress of one unit in the
$x_1$--$x_3$ plane, $\nu=1/3$. Positive and negative sectors alternate.}
\label{si:fig:angular}
\end{figure}

This section shows how the stress of one unit varies with direction and
distance, and how the stress of a row of units approaches that of a
dislocation. The purpose is to show why one unit acts mainly on its near
neighbours, while a slipped patch formed by many units produces the
long-range stress of a dislocation. Symbolic differentiation of the Kelvin double-couple
displacement gives the angular factor of the main text; in the
$x_1$--$x_3$ plane
\[
 \mathcal T_{13}=\frac{1+\nu}{2}-\frac{15}{8}\sin^2(2\theta).
\]
At $\nu=1/3$ it is $2/3$ along the axes and $-29/24$ along the diagonals,
with nodal directions in between (Figure~\ref{si:fig:angular}).

\begin{figure}[!htbp]
\centering
\includegraphics[width=.55\linewidth]{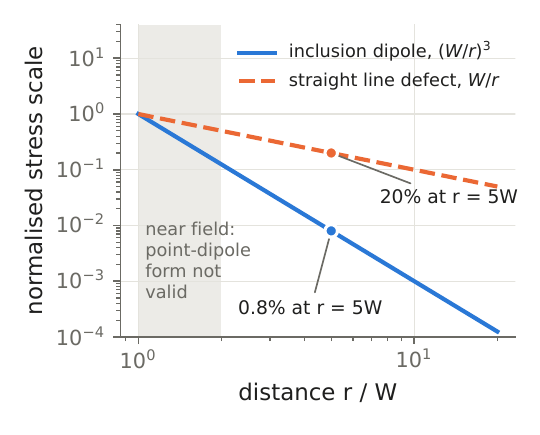}
\caption{Radial decay of the leading stress of one unit ($\propto r^{-3}$)
and of a straight dislocation ($\propto r^{-1}$), normalised at $r=W$.}
\label{si:fig:decay}
\end{figure}

Figure~\ref{si:fig:decay} compares how the stress of one unit and of a
straight dislocation fall with distance, each normalised at $r=W$. At
$r=5W$ they are $0.8\%$ and $20\%$ of their values at $r=W$; one unit
acts mainly on its near neighbours. (Close to $r\simeq W$ the full
finite-inclusion solution replaces the point-dipole form.)

\begin{figure}[!htbp]
\centering
\includegraphics[width=.6\linewidth]{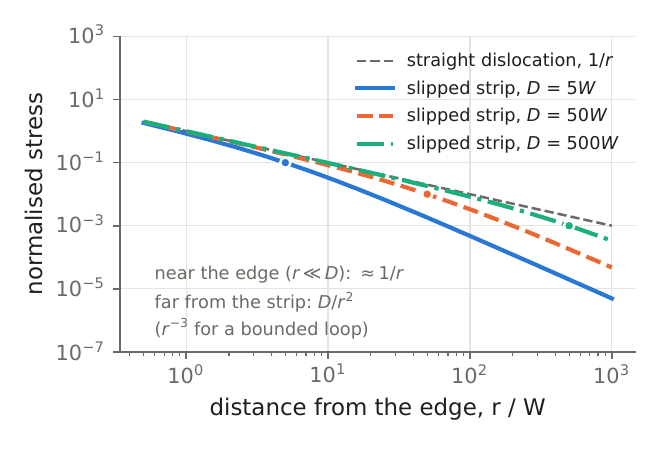}
\caption{Stress ahead of the edge of a slipped strip of width $D=5W$,
$50W$ and $500W$, compared with a straight dislocation. Markers are at
$r=D$. The $1/r$ range of a dislocation extends with $D$.}
\label{si:fig:strip}
\end{figure}

The question is whether an accumulation of units in a row results in the
emission of a line defect or a partial dislocation. When the slip of the
row has accumulated into a slipped patch, the edge of the patch carries a
dislocation whose Burgers vector is the accumulated slip.
Figure~\ref{si:fig:strip} shows the stress ahead of the edge of a slipped
strip of width $D$ in plane strain, equivalent to a pair of edge
dislocations of opposite sign, $\sigma\propto1/r-1/(r+D)$. Near the edge,
$r\ll D$, the stress follows the $1/r$ law of a dislocation; far away it
falls as $D/r^2$, and as $r^{-3}$ for a bounded patch, which is a loop.
The range of dislocation-like behaviour therefore grows with the size of
the slipped region. Emission into the grain additionally requires the
stress at the boundary to drive the edge forward, which is governed by
the kinetics of Section~10 of the main text; the number of units needed
is given in Section~6.4 of the main text (about four for a full and two
to three for a Shockley partial dislocation).

The slip profile of a row (main text, Section~6) is checked as follows;
the check confirms that the proposals P1 and P2 give the
Peierls--Nabarro form.
The density $p_W=W/[2\pi(x^2+y^2+W^2)^{3/2}]$ of proposal P1 is
normalised and projects onto the Lorentzian; its Hilbert transform gives
the stress of Equation~(51) of the main text. Figure~3 of the main text shows
why proposal P2 is needed: only coincident units give $N$ times the
profile of one unit.

\section{Flow law in transition state theory}
\begin{figure}[!htbp]
\centering
\includegraphics[width=.9\linewidth]{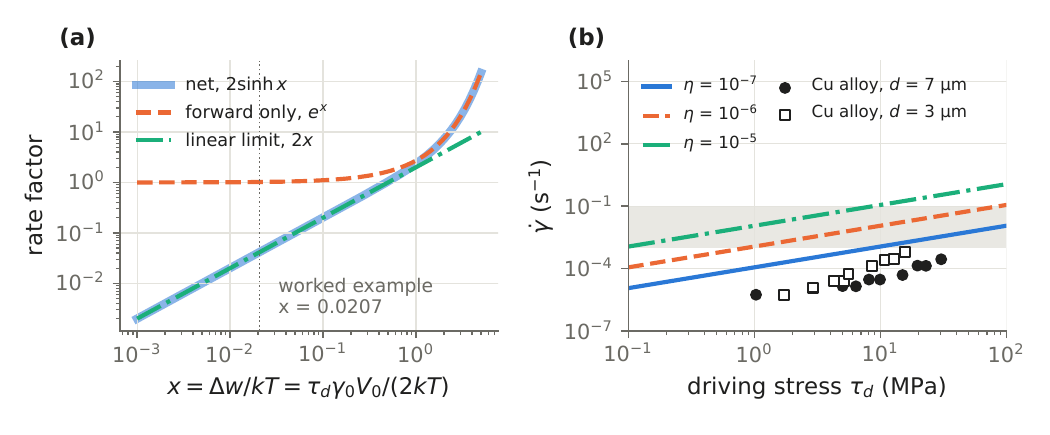}
\caption{(a) Net forward--reverse rate factor $2\sinh x$, forward-only
factor $e^{x}$ and linear limit $2x$, with $x=\tau_d\g V_0/(2kT)$.
(b) Shear rate from the transition-state rate equation for $T=773$~K,
$d=10~\mu$m, $W=0.75$~nm, $\g=0.10$, $F^\ddagger=0.40512$~eV and
$\nu_0=kT/h$, for active fractions $\eta=10^{-7}$, $10^{-6}$ and
$10^{-5}$; $\nu_0=1.61\times10^{13}$~s$^{-1}$ at 773~K. The shaded band
is the typical superplastic range $10^{-3}$--$10^{-1}$~s$^{-1}$. Symbols:
measured shear rates of a Cu-based superplastic alloy at 773~K with
$d=7$ and $3~\mu$m~\citep{paddivinski2024}, against the applied shear
stress minus the fitted threshold of that paper. The curves use the
illustrative unit of Section~S1 and are not fitted to these points; the
computation is for Cu, whereas the experimental data belong to a Cu
alloy. In (a) the net factor $2\sinh x$ (thick light-blue line)
coincides with the linear limit $2x$ for $x\lesssim0.3$ and with the
forward-only factor $e^{x}$ for $x\gtrsim3$, so that the dashed curves lie
on top of it in those ranges.}
\label{si:fig:kinetics}
\end{figure}

This section evaluates the flow law of the main text at a single boundary
and asks whether it reaches measured superplastic strain rates; the
result fixes the fraction of boundary sites that must be active. The rate equation is that of
transition state theory with a symmetric barrier, as in the
oblate-spheroid model, which applies at the elevated temperatures of
superplastic flow. For $T=773$~K, $d=10~\mu$m, $\tau_d=5$~MPa and the
unit of Section~S1, $\Delta w=0.001378707$~eV and
$\Delta w/kT=0.02069758$. With $F^\ddagger=E_{\rm el}$,
\[
 \dot\gamma=(7.0916224\times10^{-10})\eta\nu_0.
\]
Transition state theory gives the attempt frequency
$\nu_0=kT/h=1.61\times10^{13}$~s$^{-1}$ at 773~K, so
$\dot\gamma=1.14\times10^{4}\,\eta$~s$^{-1}$. A rate of $0.1$~s$^{-1}$
at $5$~MPa therefore requires an active fraction
$\eta=8.8\times10^{-6}$, and $10^{-3}$~s$^{-1}$ requires
$8.8\times10^{-8}$. With this reasonable assumption, that one boundary
site in $10^5$--$10^7$ is active at a time, the model reproduces the
superplastic strain-rate range (Figure~\ref{si:fig:kinetics}b). The active
fraction belongs to the mesoscopic description and is an input here.
This small fraction reflects the low illustrative barrier of $0.405$~eV;
with the fitted strains in the tensor convention the unit energy is
$1.6$--$1.8$~eV and no reduced active fraction is needed (Appendix A of the
main text).
Because $\Delta w/kT=0.0207$ is small, the rate is linear in $\tau_d$ to
within $0.007\%$ in this example (panel a); the $\sinh$ form matters at
higher stresses.

\begin{figure}[!htbp]
\centering
\includegraphics[width=.9\linewidth]{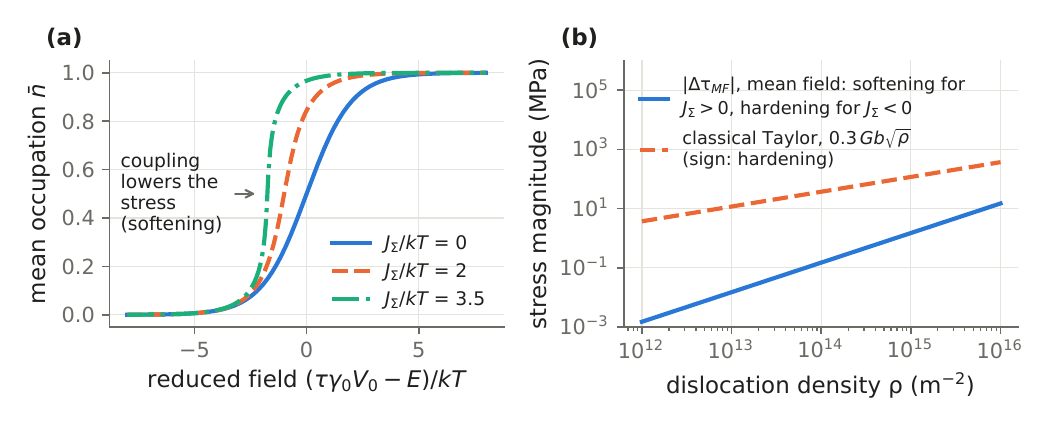}
\caption{(a) Mean occupation $\bar n$ versus reduced field for coupling
$J_\Sigma/kT=0$, 2 and 3.5. (b) Size of the mean-field stress change,
softening for $J_\Sigma>0$ and hardening for $J_\Sigma<0$, compared with
Taylor hardening for Cu-like inputs.}
\label{si:fig:meanfield}
\end{figure}

Units interact through their stress fields. Figure~\ref{si:fig:meanfield}
shows how this interaction changes the stress needed to activate them,
and what is needed for it to cause hardening rather than softening. Panel (a) solves the mean-field equation of the main text:
positive coupling moves activation to lower stress (softening). Panel
(b) evaluates the size of the mean-field stress change,
$|\Delta\tau_{\rm MF}|=z\g G\pi W^2\rho/[6(1-\nu)]$ for $z=6$,
$\g=0.10$, $W=b=0.2556$~nm and $G=48.3$~GPa, beside the Taylor scale
$0.3Gb\sqrt\rho$. Hardening of this magnitude follows if (a) neighbouring
units sit mainly in each other's negative stress sectors, so that the
net coupling is negative, and (b) the occupation is identified with the
dislocation density, $\bar n=\pi W^2\rho$. The Taylor dependence on
$\sqrt\rho$ follows if, in addition, (c) the spacing of interacting units
scales as $\rho^{-1/2}$, as in obstacle-controlled hardening.

\section{Dislocation-related estimates with experimental inputs}
\label{si:estimates}
\begin{figure}[!htbp]
\centering
\includegraphics[width=\linewidth]{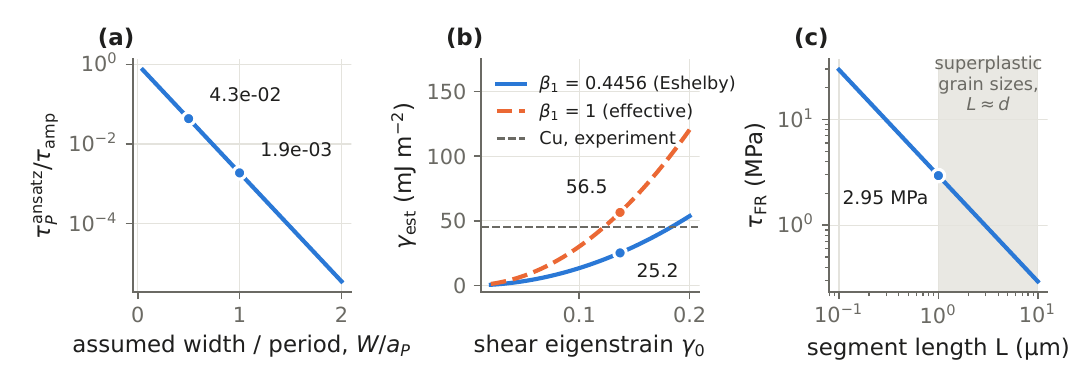}
\caption{Estimates with $G=48.3$~GPa, $b=W=0.2556$~nm and $\nu=1/3$.
(a) Peierls reduction factor versus width-to-period ratio.
(b) Stacking-fault estimate versus $\g$ for the Eshelby and an
effective unit coefficient; dashed line, measured value (45~mJ~m$^{-2}$) for
Cu~\citep{gallagher1970}. (c) Frank--Read stress versus segment length;
shaded, the grain sizes of superplastic alloys taken as $L$.}
\label{si:fig:estimates}
\end{figure}

This section shows the three estimates of Section~7 of the main text and
how material-dependent experimental inputs enter them
(Figure~\ref{si:fig:estimates}). These derivations, of the
Peierls--Nabarro and Peierls stresses, the stacking-fault energy and
Frank--Read source action, are presented to show how important relations
of dislocation theory can be obtained from the Eshelby analysis of a
single inclusion. They are not part of the argument for superplastic
flow.
(a) The Peierls stress is reduced from the lattice-resistance amplitude
by $e^{-2\pi W/a_P}$, which is $4.3\times10^{-2}$ for $W/a_P=0.5$ and
$1.9\times10^{-3}$ for $W/a_P=1$; the amplitude and period are taken from
the crystal. (b) The stacking-fault estimate for Cu is $25.2$ or
$56.5$~mJ\,m$^{-2}$ at $\g=0.137$, for $\beta_1=0.4456$ or $1$. These
bracket the measured value of about $45$~mJ\,m$^{-2}$
\citep{gallagher1970}; used as an input, the measured value gives an
effective $\beta_1\approx0.80$. (c) The Frank--Read stress is
$\tau_{\rm FR}=Gb/[2\pi(1-\nu)L]$. Taking the pinned length equal to the
grain size, $L\approx d=1$--$10~\mu$m, gives $2.95$--$0.29$~MPa. This
links Frank--Read source action to the present model through the
boundary segment from which the dislocation is emitted, which is useful
when the theory of mesoscopic superplastic flow is derived from the
oblate-spheroid model (Section~12 of the main text).

\begin{figure}[!htbp]
\centering
\includegraphics[width=.55\linewidth]{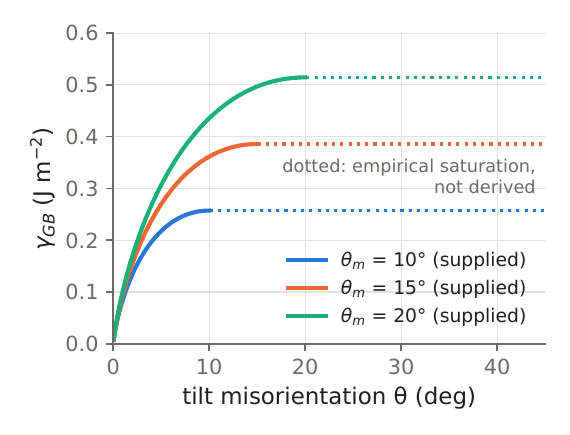}
\caption{Read--Shockley tilt-boundary energy for three crossover angles
(solid) with the usual constant continuation above them (dotted).}
\label{si:fig:rs}
\end{figure}

The energy of the boundary in which the units sit enters the mesoscopic
analyses. Figure~\ref{si:fig:rs} shows the Read--Shockley boundary energy with
$E_0=Gb/[4\pi(1-\nu)]=1.474$~J\,m$^{-2}$ for the same inputs and three
crossover angles $\theta_m$, which are experimental inputs. The energy
at $\theta_m$ is $E_0\theta_m$: $0.257$, $0.386$ and
$0.514$~J\,m$^{-2}$ for $10^\circ$, $15^\circ$ and $20^\circ$.

\section{Correspondence with the boundary-sliding models}
This section sets the present single-boundary results beside the
mesoscopic analyses, to show which quantities the unit-level analysis
reproduces and which belong only to the mesoscopic scale. It compares
them with the analyses of \citet{pad-part1}, \citet{sripathi2014} and
\citet{harisankar2025}, and states for each quantity whether it is
verified, estimated with stated inputs, not yet verified, or belongs to
the mesoscopic scale (Table~\ref{si:tab:status}).

The 2025 paper takes $W=2.5a_0$, with $a_0$ an atomic diameter, and a
shear coefficient $\beta_1=0.944(1.590-\nu)/(1-\nu)$, equal to $1.779$ at
$\nu=1/3$. In that model $\g$ is a tensor shear strain: its relation to
the uniaxial strain, $\eps=\g/\sqrt3$, follows from the tensorial von Mises
treatment. Written for the tensor shear strain, the present coefficient is
$4\times0.4456=1.782$, within $0.2\%$ of that value, so the two treatments
agree. With this convention the present coefficients reproduce the fitted
activation energies of a Cu-based superplastic alloy to within about $3\%$
(Appendix A of the main text). The 2025 fits also
impose $\eps=\g/\sqrt3$, so the fitted dilatation follows from the fitted
shear strain. The 41 rows of Section~S6 are system summaries drawn from
the 146 material conditions analysed in that paper.

\begin{table}[!htbp]
\centering
\small
\caption{Correspondence between the present analysis and the
mesoscopic boundary-sliding models.}
\label{si:tab:status}
\begin{tabular}{p{3.0cm}p{3.1cm}p{3.6cm}p{4.0cm}}
\toprule
Quantity & Present analysis & Boundary-sliding models & Status\\
\midrule
Shear coefficient $\beta_1$ & $4\times0.4456=1.782$ (tensor strain) & 1.779 & Verified: within 0.2\%\\
Dilatational coefficient $\beta_2$ & 0.8889 & From Eshelby's relation for $\beta_2$ & Derived here; material-dependent through $\nu$, linked to $\g$ by $\eps=\g/\sqrt3$\\
Unit radius $W$ & $2.5b$ & $2.5a_0$ & Common input\\
Strains $\g$ (tensor), $\eps$ & 0.05, 0.05 & fitted means 0.085, 0.049 & Shear strain below the fitted mean; dilatation within 1.5\%\\
Activation energy & $E_{\rm el}=0.405$~eV & 0.38~eV & Estimated: 6.6\% difference\\
Rate law & $\sinh$, transition state theory & $\sinh$, transition state theory & Verified: same form\\
Strain rate & $1.14\times10^4\,\eta$~s$^{-1}$ & measured & Estimated with input $\eta$\\
Units per dislocation & $N_c=4$ (full), 2--3 (partial) & --- & Not yet verified\\
Measured $Q$ & $Q/E_{\rm el}=4.05\pm1.41$ & fitted & Not yet verified\\
Boundary energy & ratio 0.91--48 & fitted & Not yet verified\\
Threshold stress $\tau_0\propto d^{-1/2}$ & Cannot be derived from a single-inclusion analysis & derived & Mesoscopic scale only\\
\bottomrule
\end{tabular}
\end{table}

The grain-size dependence of the threshold stress belongs to the
mesoscopic scale: it arises from the boundary geometry and the number of
units across a grain, which a single-boundary analysis does not contain.
It would follow from the present treatment only after the units are
summed over a network of boundaries, as in the mesoscopic model.

\section{Comparison for 41 superplastic systems}
\label{si:numerics}
This section tests the oblate-spheroid unit against data: it compares
the unit with the parameters that \citet{harisankar2025} fitted to superplastic flow data for 41 metallic
alloys, ceramics and other systems. Table~\ref{si:tab:hp} lists, for each
system, the fitted shear strain $\g$, dilatational strain $\eps$,
activation energy $Q$ and boundary energy $\gamma_B^{\rm fit}$ from that
paper, together with the authors' elastic estimates of the activation
energy $E_{\rm el}$ (eV) and boundary energy $\gamma_B^{\rm est}$
(J\,m$^{-2}$) for the same system. The last column is
$\gamma_B^{\rm fit}/\gamma_B^{\rm est}$, recomputed from the rounded
values shown. For glasses, $\gamma_B^{\rm fit}$ is a fitted interface
parameter.

\begingroup
\small
\setlength{\tabcolsep}{3pt}
\begin{longtable}{rp{3.4cm}rrrrrrr}
\caption{Rounded system summaries with recomputed ratios; author
estimates are not independently validated predictions.}
\label{si:tab:hp}\\
\toprule
\# & System & $\g$ & $\eps$ & $E_{\rm el}$ & $Q$ &
$\gamma_B^{\rm fit}$ & $\gamma_B^{\rm est}$ & ratio\\
\midrule
\endfirsthead
\multicolumn{9}{l}{Table~\ref{si:tab:hp} (continued)}\\
\toprule
\# & System & $\g$ & $\eps$ & $E_{\rm el}$ & $Q$ &
$\gamma_B^{\rm fit}$ & $\gamma_B^{\rm est}$ & ratio\\
\midrule
\endhead
\bottomrule
\endfoot
1 & Zn--22Al (2.5\,$\mu$m) & 0.059 & 0.034 & 0.195 & 0.609 & 1.29 & 0.29 & 4.45\\
2 & Zn--22Al (0.9\,$\mu$m) & 0.061 & 0.035 & 0.200 & 0.811 & 1.31 & 0.28 & 4.68\\
3 & Al--33Cu--0.4Zr & 0.066 & 0.038 & 0.263 & 1.449 & 1.61 & 0.31 & 5.19\\
4 & Al--MgScMn (3\,$\mu$m) & 0.089 & 0.051 & 0.264 & 1.310 & 1.57 & 0.17 & 9.24\\
5 & Al--MgScMn (1\,$\mu$m) & 0.083 & 0.048 & 0.260 & 1.274 & 0.92 & 0.19 & 4.84\\
6 & Al--3Mg--0.2Sc & 0.086 & 0.050 & 0.269 & 1.147 & 1.46 & 0.19 & 7.68\\
7 & Al--8.9Zn--2.6Mg--Sc & 0.080 & 0.046 & 0.267 & 1.299 & 1.47 & 0.21 & 7.00\\
8 & Al--5Mg--Mn--Sc (24\,$\mu$m) & 0.090 & 0.052 & 0.266 & 1.370 & 0.87 & 0.17 & 5.12\\
9 & Al--17Si--Fe--Mg--Cu--Ni & 0.089 & 0.051 & 0.306 & 1.126 & 0.88 & 0.20 & 4.40\\
10 & Mg--6Zn--0.8Zr & 0.080 & 0.046 & 0.264 & 1.232 & 1.54 & 0.16 & 9.62\\
11 & Mg--4Y--0.7Zr--Nd & 0.087 & 0.050 & 0.263 & 1.265 & 1.35 & 0.13 & 10.38\\
12 & Mg--5.8Zn--1Y--Zr & 0.088 & 0.051 & 0.274 & 1.088 & 1.37 & 0.13 & 10.54\\
13 & Ti--6Al--4V & 0.116 & 0.067 & 0.585 & 2.491 & 1.14 & 0.20 & 5.70\\
14 & Cu--2.8Al--Si--Co (7\,$\mu$m) & 0.096 & 0.056 & 0.394 & 1.499 & 1.48 & 0.27 & 5.48\\
15 & Cu--2.8Al--Si--Co (3\,$\mu$m) & 0.095 & 0.055 & 0.392 & 1.504 & 0.99 & 0.28 & 3.54\\
16 & IN836 (Cu--Ni--Zn) & 0.081 & 0.047 & 0.290 & 1.272 & 1.34 & 0.27 & 4.96\\
17 & Ti--43Al & 0.144 & 0.088 & 0.505 & 2.939 & 1.45 & 0.10 & 14.50\\
18 & Ti--48Al & 0.138 & 0.080 & 0.461 & 2.962 & 1.20 & 0.11 & 10.91\\
19 & Ti--46.2Al--2.2Cr & 0.125 & 0.072 & 0.460 & 2.723 & 1.07 & 0.13 & 8.23\\
20 & Co$_3$Ti & 0.096 & 0.055 & 0.414 & 2.824 & 0.86 & 0.28 & 3.07\\
21 & Ni$_3$Si & 0.092 & 0.053 & 0.465 & 2.288 & 0.87 & 0.34 & 2.56\\
22 & ZrO$_2$ & 0.082 & 0.047 & 1.256 & 3.481 & 1.65 & 0.57 & 2.89\\
23 & ZrO$_2$--3Y & 0.088 & 0.051 & 1.261 & 3.790 & 0.85 & 0.49 & 1.73\\
24 & ZrO$_2$--4Y & 0.089 & 0.051 & 1.253 & 3.909 & 1.57 & 0.48 & 3.27\\
25 & Al$_2$O$_3$--ZrO$_2$--mullite & 0.067 & 0.039 & 1.130 & 3.473 & 0.86 & 0.75 & 1.15\\
26 & Al$_2$O$_3$--NiAl$_2$O$_4$--ZrO$_2$ & 0.064 & 0.037 & 1.244 & 3.690 & 1.35 & 0.91 & 1.48\\
27 & 6061 / 20\% SiC & 0.071 & 0.041 & 0.439 & 1.556 & 1.05 & 0.45 & 2.33\\
28 & 7075 / 20\% SiC & 0.069 & 0.040 & 0.421 & 1.537 & 0.97 & 0.45 & 2.16\\
29 & Zr$_{65}$ BMG & 0.074 & 0.043 & 0.253 & 1.438 & 1.35 & 0.26 & 5.19\\
30 & Zr$_{52.5}$ BMG & 0.075 & 0.043 & 0.268 & 1.518 & 0.78 & 0.27 & 2.89\\
31 & La$_{55}$ BMG & 0.059 & 0.034 & 0.151 & 1.015 & 1.60 & 0.21 & 7.62\\
32 & La$_{60}$ BMG & 0.073 & 0.042 & 0.213 & 0.963 & 0.82 & 0.19 & 4.32\\
33 & Fe$_{72}$ BMG & 0.079 & 0.046 & 0.346 & 1.452 & 0.78 & 0.35 & 2.23\\
34 & Limestone & 0.073 & 0.042 & 1.045 & 2.554 & 0.57 & 0.29 & 1.97\\
35 & Anorthite--Diopside (dry) & 0.062 & 0.036 & 1.601 & 3.117 & 1.12 & 0.61 & 1.84\\
36 & Anorthite--Diopside (wet) & 0.062 & 0.036 & 1.624 & 2.697 & 1.27 & 0.62 & 2.05\\
37 & Ice (10\,$\mu$m) & 0.115 & 0.067 & 0.169 & 0.332 & 0.79 & 0.03 & 26.33\\
38 & Ice (1700\,$\mu$m) & 0.117 & 0.068 & 0.173 & 0.408 & 1.44 & 0.03 & 48.00\\
39 & Si$_3$N$_4$ & 0.085 & 0.049 & 2.765 & 3.168 & 1.05 & 1.15 & 0.91\\
40 & Zirconia--Alumina & 0.069 & 0.040 & 1.311 & 3.166 & 0.94 & 0.83 & 1.13\\
41 & Zirconia--Spinel & 0.075 & 0.043 & 1.180 & 3.434 & 1.62 & 0.64 & 2.53\\
\end{longtable}
\endgroup

\begin{table}[!htbp]
\centering
\caption{Statistics calculated from the rounded entries above. Standard
deviations use $n-1$ in the denominator. Relative differences use the
transcribed mean as denominator.}
\label{si:tab:stats}
\begin{tabular}{lr}
\toprule
Quantity & Recomputed value\\
\midrule
Mean shear strain $\pm$ sample s.d. & $0.08510\pm0.02050$\\
Mean dilatational strain $\pm$ sample s.d. & $0.04927\pm0.01225$\\
Relative difference: $\g=0.10$ & $+17.51\%$\\
Relative difference: $\g=0.12$ & $+41.01\%$\\
Relative difference: $\eps=0.05$ & $+1.49\%$\\
$Q/E_{\rm el}$ mean $\pm$ sample s.d. & $4.047\pm1.411$\\
$\gamma_B^{\rm fit}/\gamma_B^{\rm est}$ median & $4.45$\\
Ratio range & $0.91$--$48.0$\\
Within factors $3$, $5$, $10$ & $15/41$, $24/41$, $35/41$\\
\bottomrule
\end{tabular}
\end{table}

\begin{figure}[!htbp]
\centering
\includegraphics[width=\linewidth]{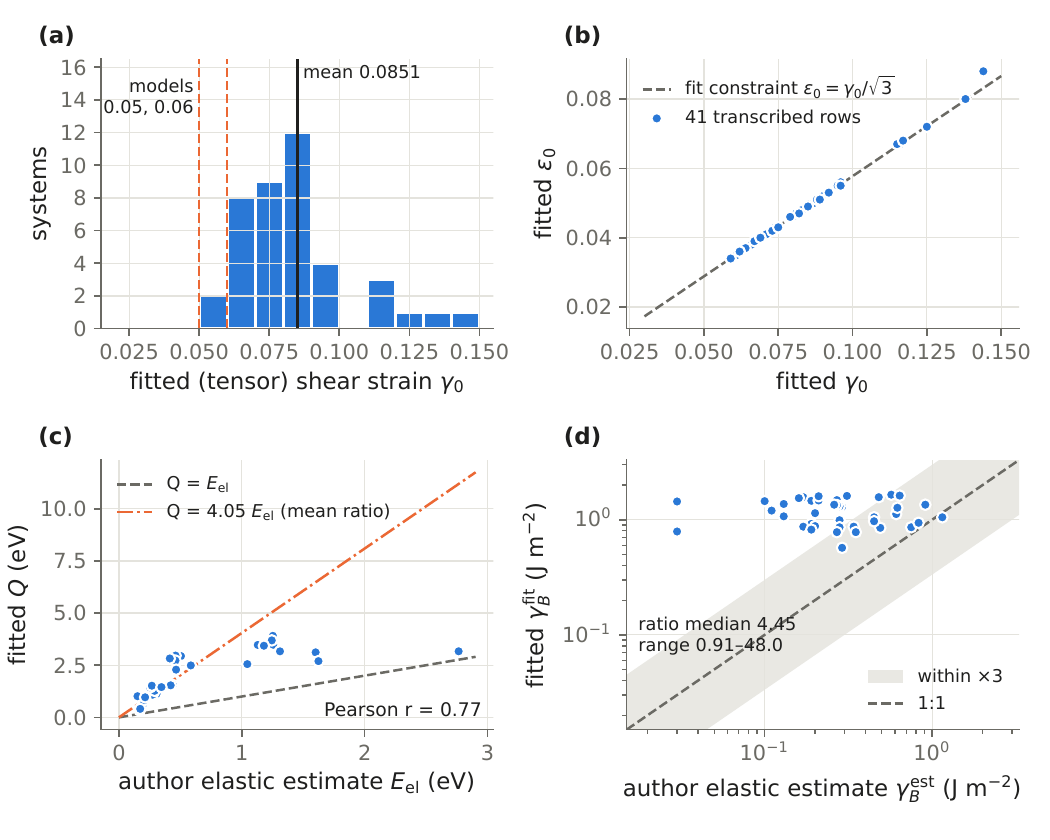}
\caption{Comparison with the fits of \citet{harisankar2025} for 41
systems. (a) Fitted (tensor) shear strain, with its mean and the illustrative
values in the same convention, 0.05 and 0.06. (b) Fitted dilatation against fitted shear strain; dashed,
the relation $\eps=\g/\sqrt3$ used in the fits. (c) Fitted activation
energy against the elastic energy of one unit. (d) Fitted against
estimated boundary energy; shaded, within a factor of 3.}
\label{si:fig:hp25}
\end{figure}

Table~\ref{si:tab:stats} summarises these entries. The comparison shows three things. First, the fitted dilatation is close
to the model value: $0.05$ is $1.5\%$ above the mean $0.049$. The fitted
shear strain is a tensor strain with mean $0.085$, i.e.\ an engineering
strain of $0.170$, so the illustrative engineering value $0.10$ (tensor
$0.05$) lies below it (Figure~\ref{si:fig:hp25}a,b). Second, the fitted activation energy scales
with the elastic energy of one unit (Pearson $r=0.77$), with
$Q\approx4E_{\rm el}$ (panel c). These estimates treat the fitted strains as
engineering strains. With the tensor convention of the boundary-sliding model, the
elastic estimate rises by a factor of $2.80$ for $\eps=\g/\sqrt3$, and the
mean ratio becomes $Q/E_{\rm el}=1.45\pm0.50$; most of the apparent
factor of four is therefore the strain convention.

The two conventions differ as follows. For simple shear by an angle
$\gamma$, the engineering shear strain is $\gamma$ itself, whereas the
tensor shear strain is the off-diagonal component of the strain tensor,
$\varepsilon_{13}=\gamma/2$. The present analysis writes the eigenstrain
as $\varepsilon^*_{13}=\g/2$, so its $\g$ is an engineering strain, and the
shear energy is $\tfrac12\beta_1G\g^2V_0$. In the boundary-sliding model
the fitted $\g$ is a tensor shear strain, so the same energy reads
$\tfrac12(4\beta_1)G\g^2V_0$; the factor $4\beta_1=1.782$ is the
coefficient compared with $1.779$ in Table~\ref{si:tab:status}. The
relation $\eps=\g/\sqrt3$ of that model links the tensor shear strain to
the uniaxial strain through the von Mises equivalence. Reading the fitted
$\g$ as a tensor strain therefore multiplies the shear energy by four;
with $\eps=\g/\sqrt3$ the total elastic energy rises by
$(4\beta_1+\beta_2/3)/(\beta_1+\beta_2/3)=2.80$.

Third, the estimated boundary energies are within a factor of
3 of the fitted values for 15 of the 41 systems and within a factor of
10 for 35 (panel d); the two ice rows are sensitive to rounding of their
small estimates.

\section{Simulation details for Appendix A}
\label{si:methods}
This section gives the settings needed to reproduce the
molecular-dynamics calculations of Appendix~A of the main text, which
illustrate the route from a single boundary to the mesoscopic scale. All simulations use the embedded-atom
potential for Cu of \citet{mishin2001} in LAMMPS~\citep{thompson2022}
(version of 22 July 2025). Structures were built with
Atomsk~\citep{hirel2015}; structural analysis used OVITO~3.12
\citep{stukowski2010}, with adaptive common-neighbour analysis and the
dislocation extraction algorithm~\citep{stukowski2012}. The runs are
listed in Table~\ref{si:tab:runs}. All inputs, starting structures, analysis
scripts and processed data are available at \url{https://github.com/albert-hzbn/oblate-spheroid-gbs}.

\begin{table}[!htbp]
\centering
\small
\caption{Simulations of Appendix~A of the main text.}
\label{si:tab:runs}
\begin{tabular}{p{4.0cm}p{4.2cm}p{5.4cm}}
\toprule
Calculation & System & Conditions\\
\midrule
Stacking-fault energy & $(111)$ slab, 36 planes & 0~K, partial and full paths\\
Grain-size series & 16 grains, $d=3$--$16$~nm, $1.9\times10^4$--$2.9\times10^6$ atoms & 300~K, $5\times10^8$~s$^{-1}$, 10\% strain; two seeds for $d\le10$~nm\\
Rate--temperature series & 16 grains, $d=5$~nm, $8.9\times10^4$ atoms & 300--700~K, $10^8$--$3\times10^9$~s$^{-1}$\\
Quasi-static shear & Three $\Sigma11$ structures and $\Sigma7$, both shear senses & 0~K, strain step 0.004, up to 0.36\\
Nudged elastic band & $\Sigma11$ bicrystal, 135{,}168 atoms & 0.71--1.08~GPa; 8 and 16 replicas\\
\bottomrule
\end{tabular}
\end{table}

\paragraph{Generalised stacking-fault energy.} A slab with $x$, $y$ along
$[11\bar2]$ and $[\bar110]$ (partial path) or $[\bar110]$ and $[\bar1\bar12]$
(full path) and $z$ along $[111]$ contained 36 $(111)$ planes, with
in-plane dimensions equal to whole multiples of the repeat distances
($6\times10$ repeats) and free surfaces normal to $z$. The upper half was
displaced rigidly in steps along the slip direction, and atoms were
relaxed along $z$ only by conjugate-gradient minimisation to a force
tolerance of $10^{-10}$~eV\,\AA$^{-1}$.

\begin{figure}[!htbp]
\centering
\includegraphics[width=\linewidth]{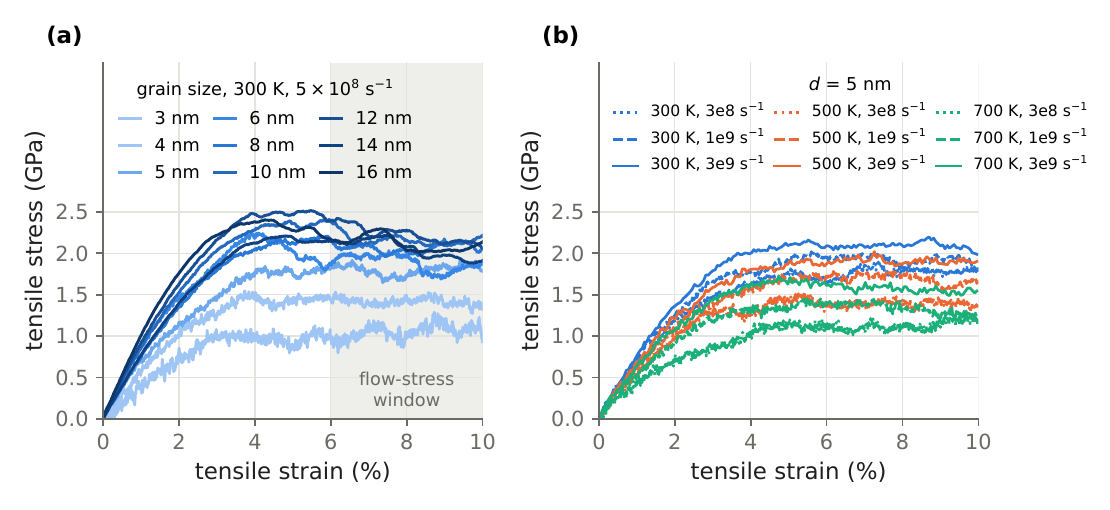}
\caption{Tensile stress--strain curves of nanocrystalline Cu. (a) Grain
sizes 3--16~nm at 300~K and $5\times10^8$~s$^{-1}$ (first seed); the shaded
band is the strain window used for the flow stress. (b) $d=5$~nm at three
temperatures and three strain rates.}
\label{si:fig:ss}
\end{figure}

\paragraph{Nanocrystalline polycrystals.} Sixteen grains with random
centres and orientations were placed in a periodic cubic box of side
$L=2.031\,d$, where $d$ is the equivalent-sphere grain diameter. Atoms
closer than $2.0$~\AA\ were removed, the structure was minimised, and it
was annealed for 20~ps at the test temperature and zero pressure
(Nos\'e--Hoover thermostat and barostat, damping times $0.1$ and $1.0$~ps,
time step 2~fs). Uniaxial tension was then applied along $x$ at constant
engineering strain rate to 10\% strain, with zero lateral stress. The flow
stress is the mean tensile stress between 6 and 10\% strain
(Figure~\ref{si:fig:ss}). Dislocation content and the fraction of
hexagonal-close-packed atoms were evaluated on the snapshot at 10\%
strain. Activation volumes follow from
$V^*=\sqrt3\,kT/(\partial\sigma/\partial\ln\dot\varepsilon)$ at each
temperature, and the apparent activation energy at fixed stress from the
slope of $\ln\dot\varepsilon$ against $1/T$.

\paragraph{Quasi-static shear of bicrystals.} The periodic
$\Sigma11(557)/(77\bar1)$ bicrystals (135{,}168 atoms, two boundaries) and
the $\Sigma7(415)$ bicrystal were sheared by increasing the $yz$ tilt in
steps of $0.004$ engineering strain, each followed by conjugate-gradient
minimisation at fixed cell to $10^{-6}$~eV\,\AA$^{-1}$. An emission event
was identified as the first step with a shear-stress drop above
$0.15$~GPa together with an increase of more than 500 atoms with
hexagonal-close-packed coordination (cut-off $3.087$~\AA); dislocation
line content before and after the step was compared by the dislocation
extraction algorithm.

\paragraph{Nudged elastic band.} Initial states were the minimised
configurations at the chosen strains. Final states were the configuration
just after the instability, mapped into the cell of the initial state by
its fractional coordinates and minimised to $10^{-8}$~eV\,\AA$^{-1}$; it
was used only if the emitted fault persisted. Climbing-image nudged
elastic band calculations~\citep{henkelman2000} applied springs of
$1.0$~eV\,\AA$^{-2}$ to all atoms, with the FIRE minimiser (time step
1~fs), 3000 ordinary and up to 4000 climbing iterations, and a replica
force tolerance of $0.05$~eV\,\AA$^{-1}$, with 8 and 16 replicas.

\paragraph{Digitised superplastic data.} The experimental points of
Figure~A1 of \citet{paddivinski2024} were digitised from the published
image after calibrating its logarithmic axes on the plot frame. Markers
were located by colour, thin model lines were removed by morphological
opening, and each marker was matched by strain rate to the tabulated
experimental value; only its stress was taken from the image. Of 55
tabulated points, 52 were matched. The lowest stress of each data set
agrees with the tabulated minimum (a tensile stress, divided by
$\sqrt3$) to within $1\%$.

\clearpage

\end{document}